\documentclass[11pt]{article}

\pdfoutput=1

\usepackage{amsmath,amssymb,mathrsfs}
\usepackage{epstopdf,graphicx}
\usepackage{caption,subcaption}
\usepackage{fullpage,setspace}
\usepackage{apacite}

\newcommand{\xpos}{\boldsymbol{x}}
\newcommand{\ypos}{\boldsymbol{y}}

\begin{document}

\title{\bf Synthetic aperture radar imaging below a random rough surface}

\author{Arnold D.~Kim and Chrysoula Tsogka}

\date{Department of Applied Mathematics, University of California,
  Merced\\ 5200 North Lake Road, Merced, CA 95343, USA}

\maketitle

\begin{abstract}
  Motivated by applications in unmanned aerial based ground
  penetrating radar for detecting buried landmines, we consider the
  problem of imaging small point like scatterers situated in a lossy
  medium below a random rough surface. Both the random rough surface
  and the absorption in the lossy medium significantly impede the
  target detection and imaging process. Using principal component
  analysis we effectively remove the reflection from the air-soil
  interface.  We then use a modification of the classical synthetic
  aperture radar imaging functional to image the targets. This imaging
  method introduces a user-defined parameter, $\delta$, which scales
  the resolution by $\sqrt{\delta}$ allowing for target localization
  with sub wavelength accuracy. Numerical results in two dimensions
  illustrate the robustness of the approach for imaging multiple
  targets. However, the depth at which targets are detectable is
  limited due to the absorption in the lossy medium.
\end{abstract}

\section{Introduction}

Landmine detection using unmanned aerial based radar is gaining
attention because it provides high resolution images while avoiding
the interaction with the object and the surrounding medium
\cite{Fernadez18, Francke21}. Those imaging systems use synthetic
aperture radar (SAR) processing to achieve high resolution imaging of
both metallic and dielectric targets. In SAR, high resolution is achieved
because the data are treated coherently along the flight path of a
single transmitter/receiver mounted on an aircraft. For landmine detection, SAR image
processing is used and the data are coherently processed along the
synthetic aperture formed by an unmanned aerial vehicle flying above
the ground over the area of interest. Other related remote sensing
applications include precision agriculture, forestry monitoring and
glaciology.

Landmine detection is a very important problem with both civilian and
military applications. It has been a subject of extreme interest and
several imaging methodologies have been proposed in the literature. We
refer to the review article \cite{daniels:2006} for an overview on the
subject and to \cite{Soldovieri2014} for a comparison between
different imaging techniques in the specific context of landmine
detection. The method we employ here is a modification
of the classical SAR processing technique. Specifically we apply to
the classical imaging functional a M\"obius transformation that
depends on a user defined parameter, $\delta$. Assuming a synthetic
aperture of length $a$, and system bandwidth $B$, we have recently
shown \cite{KimTsogka-Tunable} that the resolution of the imaging
method in cross-range (the direction parallel to the synthetic
aperture) is $\sqrt{\delta} \lambda L/a$ and the range (direction
orthogonal to cross-range) resolution is $\sqrt{\delta}c/B$ with $c$
the speed of the waves, $\lambda$ the central wavelength and $L$ the
distance of propagation. We have also carried out a resolution
analysis of this method for imaging in a lossy medium \cite{KT-Lossy}
where we have shown that one should not use the absorption in the
medium even if it is known. Although, absorption does not affect
significantly the resolution of the imaging method, it does affect the
target detectability. Specifically, if $z$ denotes the depth of the
target below the air-soil interface, the product $\beta z$ corresponds
to the absorption length scale of the problem with $\beta$ denoting
the loss tangent, that is the ratio of the imaginary part over the
real part of the relative dielectric constant. For targets buried deep
so that $\beta z \gg 1$ measurements become too small to detect
targets, especially if the data are corrupted by additive measurement
noise as is often the case in practical applications.

For a sufficiently long flight path, the air-soil interface is most
likely not uniformly flat. Moreover, height fluctuations in this
interface cannot be known with certainty. For this reason we model
this interface using a random rough surface. It then becomes crucially
important for a subsurface imaging method to be robust to those
uncertainties in the interface. Additionally, there may be multiple
interactions between scattering by subsurface targets and the random
rough surface \cite{long2010scattering}. Here, we assume only one
interaction between the random rough surface and the subsurface target
since that has been shown to be sufficiently accurate for targets
buried in a lossy medium \cite{El-Shenawee02}.

We model the height of the air-soil interface $h(x)$ using a
Gaussian-correlated random process that is characterized by the RMS
height, $h_{\text{RMS}}$ and the correlation length, $\ell$. We
consider here that the RMS height is small with respect to the
correlation length which is of the order of the central wavelength
while the aperture is large compared to both. In this regime,
multiple-scattering effects are important and enhanced backscattering
is observed. Enhanced backscattering is a multiple scattering
phenomenon in which a well-defined peak in the retro-reflected
direction is observed \cite{Maradudin91, Ishimaru91,
  Maradudin2007}. Imaging in media with random rough surfaces is a new
paradigm for imaging in random media and requires different methods
than the ones developed for volumetric scattering \cite{BGPT-rtt} or
imaging in random waveguides \cite{BGT-waveg}. The key difference here
is that randomness is isolated only at the interface separating the
two media. Even though waves multiply scatter on the rough surface,
they also scatter away from the rough surface. Consequently, there is
no dominant cumulative diffusion phenomenon due to this kind of
randomness.

For the synthetic aperture setup the measurements are exactly in the
retro-reflected direction so the data have uniform power at each
spatial location along the flight path.  To remove the strong
reflection introduced by the ground-air interface we use PCA or more
precisely the singular value decomposition (SVD) of the data matrix.
Principal component analysis (PCA) has been proposed as a method for
removing ground bounce signals in ~\cite{tjora2004evaluation}. For a
flat surface the ground bounce can be removed from the data by taking
out the contribution corresponding to the first singular value. Here
we see that due to multiple scattering to remove the reflection from
the random interface contributions corresponding to the first few
singular values should be taken out from the data. This SVD based
approach for ground bounce removal is advantageous because it does not
require any {\it a priori} information about the media, including the
exact location of the interface.

Our imaging method requires computing Green's function for a medium
composed of adjacent half spaces. This Green's function is represented
as a Fourier integral of a highly oscillatory function. Accurately
computing such integrals is quite challenging and several approaches
have been proposed to this effect \cite{cai2002algorithmic,
  o2014efficient, bruno2016windowed}. The approach we follow here is
similar to the method presented by Barnett and
Greengard~\cite{barnett2011new}, where we integrate on a deformed
contour in the complex plane to avoid branch points.

The remainder of the paper is as follows.  In Section \ref{sec:SAR} we
present the synthetic aperture radar setup. In Section \ref{sec:rough}
our model for the rough surface is described as well as the integral
equations formulation for computing the solution to the forward
problem. The algorithm for computing the measurements is then
explained in Section \ref{sec:measurements}.  The solution of the
inverse scattering problem entails two steps.  The first step that
uses the singular value decomposition of the data matrix to remove the
ground bounce is presented in Section \ref{sec:ground-bounce}. The
second step consists in reconstructing an image using the modified
synthetic aperture imaging algorithm and is explained in Section
\ref{sec:KM}. We present numerical results in two dimensions that
illustrate the effectiveness of the imaging method in Section
\ref{sec:numerics}.  We finish with our conclusions in Section
\ref{sec:conclusions}.

\section{SAR imaging}
\label{sec:SAR}

Here we describe the SAR imaging system for the problem to be studied.
We limit our computations to the two-dimensional $xz$-plane to
simplify the simulations. However, the imaging method we describe
easily extends to three-dimensional problems.

\begin{figure}[htbp]
  \centering
    \includegraphics[width=0.6\linewidth]{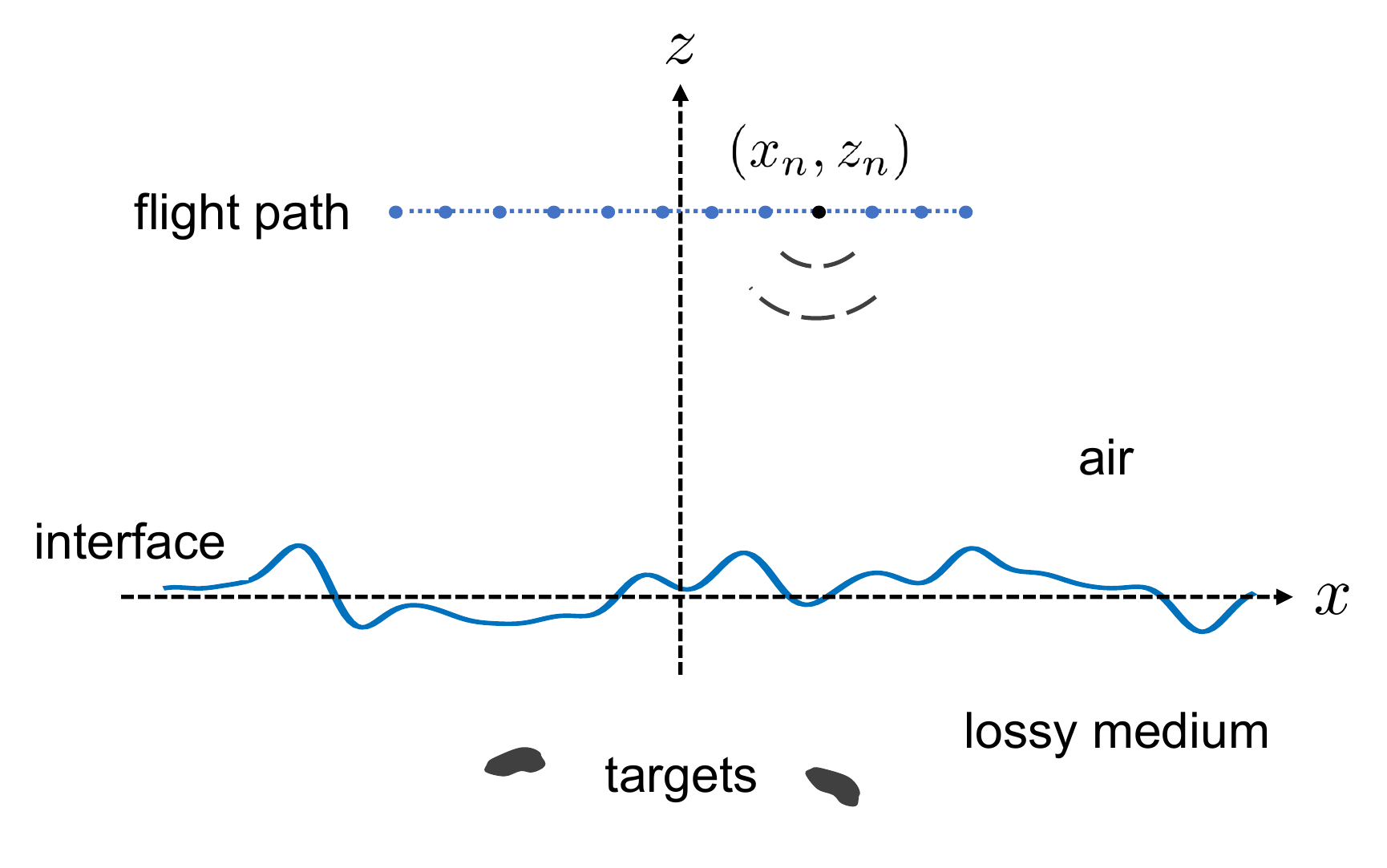}
    \caption{A sketch of the subsurface synthetic aperture imaging
      system. A platform moves along a prescribed flight path
      producing a synthetic aperture above an interface separating air
      from a lossy medium. The platform emits a signal and records the
      echoes including ground bounce signals due to reflections by the
      interface and scattered signals by the targets. The objective
      for the imaging problem is to identify and locate the subsurface
      targets.}
  \label{fig:SAR-problem}
\end{figure}

Consider a platform moving along a prescribed flight path. At fixed
locations along the flight path: $\xpos_{n} = (x_{n}, z_{n})$ for
$n = 1, \dots, N$, the platform emits a multi-frequency signal that
propagates down to an interface that separates the air where the
platform is moving from a lossy medium below the interface. See
Fig.~\ref{fig:SAR-problem} for a sketch of this imaging system. Let
$\omega_{m}$ for $m = 1, \dots, M$ denote the set of frequencies used
for emitting and recording signals. We apply the start-stop
approximation here in which we neglect the motion of the platform and
targets in comparison to the emitting and recording of signals.  The
complete set of measurements corresponds to the suite of experiments
conducted at each location on the path. 

For this problem, the signal emitted from the platform propagates down
to the interface. Part of the signal is reflected by the interface
which is called the ground bounce signal. The portion of that ground
bounce signal that reaches the platform is recorded. Another part of
the signal is transmitted across the interface and is incident on the
subsurface targets which then scatter that signal. Since the medium
below the interface is lossy, the power in the signals incident on and
scattered by the targets is attenuated. A portion of that attenuated
scattered signal is transmitted across the interface and propagates up
to the platform where it is also recorded. Measurements are therefore
comprised of ground bounce and scattered signals reaching the
platform.

Using these measurements we seek to solve the inverse scattering
problem that identifies and locates targets in the lossy medium below
the interface. The medium above the interface is uniform and lossless
and we assume that it is known. The medium below is also uniform, but
lossy, so it has a complex relative dielectric permittivity. We assume
we know the real part of the relative dielectric permittivity, but not
its imaginary part corresponding to the absorption in the medium.
Finally, the interface between the two media is unknown, but we assume
that we know its mean, which is constant.

There are several key challenges to consider for this
problem. Measurements include ground bounce and scattered signals. The
ground bounce signals have more power than the scattered signals, but
do not contain information about the targets. Thus, one needs an
effective method to remove the ground bounce from
measurements. Because the interface is uncertain, it is important to
remove these ground bounce signals without requiring explicit
knowledge of the interface location. Once that issue can be adequately
addressed, we then require high-resolution images of the targets in an
unknown, lossy medium obtained through solution of the inverse
scattering problem. The absorption in the medium will limit the depth
at which one can reliably solve the inverse scattering
problem. However, we are interested in identifying targets that are
located superficially below the interface, so the penetration depths
needed for this problem are not too prohibitive. In addition,
measurements are corrupted by additive measurement noise. Another
noteworthy issue is that removal of the ground bounce signal from
measurements will effectively increase the relative amount of noise in
what remains which will limit the values of the signal-to-noise ratio
(SNR) for which imaging will be effective.

\section{Rough surface scattering}
\label{sec:rough}

We model uncertainty in the interface separating the two media using
random rough surfaces. In particular, we consider Gaussian-correlated
random surfaces that are characterized by the RMS height,
$h_{\text{RMS}}$ and the correlation length, $\ell$. In what follows,
we give the integral equation formulation for computing reflection and
transmission of signals across one realization of a random rough
surface.

Let $z = h(x)$ for $-\infty < x < \infty$ denote one realization of
the random rough surface separating two different media. The medium in
$z > h(x)$ is uniform and lossless. The medium in $z < h(x)$ is also
uniform, but lossy with relative dielectric constant
$\epsilon_{r} (1 + \mathrm{i} \beta )$ with $\epsilon_{r}$ denoting
the real part of the relative dielectric constant and $\beta \ge 0$
denoting the loss tangent (ratio of the imaginary part over the real
part of the relative dielectric constant). We consider two problems in
which a point source is either above or below the interface.  In what
follows we assume that the total field and its normal derivative are
continuous on $z = h(x)$ and that those fields satisfy appropriate
out-going conditions as $z \to \pm \infty$.

\subsection{Integral equations formulation}

Suppose a point source is located at $(x_{0},z_{0})$ with $z_{0} >
h(x_{0})$. Using Green's second identity, we write
\begin{equation}
  u(x,z) = G_{0}(x,z;x_{0},z_{0})
  + \mathscr{D}_{0}[U](x,z) - \mathscr{S}_{0}[V](x,z), \quad z > h(x),
  \label{eq:uBIE0}
\end{equation}
with
\begin{equation*}
  \mathscr{D}_{0}[U](x,z) = \int_{-\infty}^{\infty} \frac{\partial
    G_{0}(x,z;\xi,h(\xi))}{\partial n} \sqrt{1 + (h'(\xi))^{2}} U(\xi)
  \mathrm{d}\xi,
\end{equation*}
and
\begin{equation*}
  \mathscr{S}_{0}[V](x,z) = \int_{-\infty}^{\infty}
  G_{0}(x,z;\xi,h(\xi)) V(\xi) \mathrm{d}\xi.
\end{equation*}
Here,
\begin{equation*}
  G_{0}(x,z;x',z') = \frac{\mathrm{i}}{4} H_{0}^{(1)}\left( k_{0}
    \sqrt{ ( x - x' )^{2} + ( z - z' )^{2}} \right),
\end{equation*}
with $k_{0} = \omega / c$ and
\begin{equation}
  \frac{\partial G_{0}(x,z;\xi,\zeta)}{\partial n} \sqrt{1 +
    (h'(\xi))^{2}} = h'(\xi) \frac{\partial
    G_{0}(x,z;\xi,\zeta)}{\partial \xi} - \frac{\partial
    G_{0}(x,z;\xi,\zeta)}{\partial \zeta}.
  \label{eq:normalD}
\end{equation}
In addition, we have
\begin{equation}
  v(x,z) = -\mathscr{D}_{1}[U](x,z) + \mathscr{S}_{1}[V](x,z), \quad z
  < h(x),
  \label{eq:vBIE0}
\end{equation}
with $\mathscr{D}_{1}$ and $\mathscr{S}_{1}$ defined the same as
$\mathscr{D}_{0}$ and $\mathscr{S}_{0}$, but with $G_{0}$ replaced
with
\begin{equation*}
  G_{1}(x,z;x',z') = \frac{\mathrm{i}}{4} H_{0}^{(1)}\left( k_{1}
    \sqrt{ ( x - x' )^{2} + ( z - z' )^{2}} \right),
\end{equation*}
and $k_{1} = k_{0} \sqrt{\epsilon_{r}( 1 + \mathrm{i} \beta )}$.  Now,
suppose a point source is located at $(x_{1},z_{1})$ with
$z_{1} < h(x_{1})$. For that case we have
\begin{equation}
  u(x,z) = \mathscr{D}_{0}[U](x,z) - \mathscr{S}_{0}[V](x,z), \quad z
  > h(x),
  \label{eq:uBIE1}
\end{equation}
and
\begin{equation}
  v(x,z) = G_{1}(x,z;x_{1},z_{1}) - \mathscr{D}_{1}[U](x,z) +
  \mathscr{S}_{1}[V](x,z), \quad z < h(x).
  \label{eq:vBIE1}
\end{equation}

The fields $u$ defined by either \eqref{eq:uBIE0} or \eqref{eq:uBIE1},
and $v$ defined by either \eqref{eq:vBIE0} or \eqref{eq:vBIE1} are
given in terms of surface fields $U(\xi)$ and $V(\xi)$. Physically,
$U(\xi) = u(\xi,h(\xi))$ is the evaluation of the field on the
interface point, $(\xi, h(\xi))$. The field $V(\xi)$ is defined in
terms of the normal derivative of $u$ according to
\begin{equation*}
  V(\xi) = \sqrt{1 + (h'(\xi))^{2}} \frac{\partial
    u(\xi,h(\xi))}{\partial n} = h'(\xi) \frac{\partial
    u(\xi,\zeta)}{\partial \xi} - \frac{\partial
    u(\xi,\zeta)}{\partial \zeta}.
\end{equation*}
These formulations given above make use of the aforementioned
assumption that both $u$ and $\partial_{n} u$ are continuous on the
interface $z = h(x)$.

The surface fields $U$ and $V$ are not yet determined. To determine
them we evaluate $u$ and $v$ in the limit as $(x,z) \to (\xi,h(\xi))$
from above and below, respectively. In that limit, the
$\mathscr{D}_{0}$ and $\mathscr{D}_{1}$ operators produce a jump and
the result is a system of boundary integral equations. For the fields
defined by \eqref{eq:uBIE0} and \eqref{eq:vBIE0}, the resulting system
is
\begin{subequations}
   \begin{align}
     \frac{1}{2} U(\xi) - \mathscr{D}_{0}[U](\xi) +
     \mathscr{S}_{0}[V](\xi) &= G_{0}(\xi,h(\xi);x_{0},z_{0}), \\
     \frac{1}{2} U(\xi) + \mathscr{D}_{1}[U](\xi) -
     \mathscr{S}_{1}[V](\xi) &= 0,
   \end{align}
   \label{eq:BIE-top}
\end{subequations}
and for the fields defined by \eqref{eq:uBIE1} and \eqref{eq:vBIE1},
the resulting system is
\begin{subequations}
   \begin{align}
     \frac{1}{2} U(\xi) - \mathscr{D}_{0}[U](\xi) +
     \mathscr{S}_{0}[V](\xi) &= 0, \\
     \frac{1}{2} U(\xi) + \mathscr{D}_{1}[U](\xi) -
     \mathscr{S}_{1}[V](\xi) &= G_{1}(\xi,h(\xi);x_{1},z_{1}).
   \end{align}
   \label{eq:BIE-bottom}
\end{subequations}
The solution of each of these systems results in the determination of $U$
and $V$ for their respective problem. Once those are determined, the
fields above and below the interface are computed through evaluation
of \eqref{eq:uBIE0} and \eqref{eq:vBIE0} when the source is above the
interface, or \eqref{eq:uBIE1} and \eqref{eq:vBIE1} when the source is
below the interface. We give the numerical method we use to solve
these systems in the Appendix.

\subsection{Enhanced backscattering}

The bistatic cross-section $\sigma(\theta_{s},\theta_{i})$ is the
fraction of power reflected in the far field by the rough surface in
direction $(\sin\theta_{s}, \cos\theta_{s})$ with $\theta_{s}$
denoting the scattered angle made with respect to the $z$-axis due to
a plane wave incident in direction $(\sin\theta_{i}, -\cos\theta_{i})$
with $\theta_{i}$ denoting the angle of incidence.  Reflection by the
random rough surface makes up an important component of measurements
in this imaging problem. Here, we use the bistatic cross-section to
characterize reflection by the rough surface over the range of
frequencies: $3.1$ GHz to $5.1$ GHz. We use the method given in
\cite[Chapter 4]{tsang2004scattering} to generate these rough surfaces
and compute the corresponding bistatic cross-sections. We then average
over several realizations of the rough surface to determine canonical
features of these rough surfaces.

\begin{figure}[t]
  \centering
  \hfill
  \includegraphics[width=0.40\linewidth]{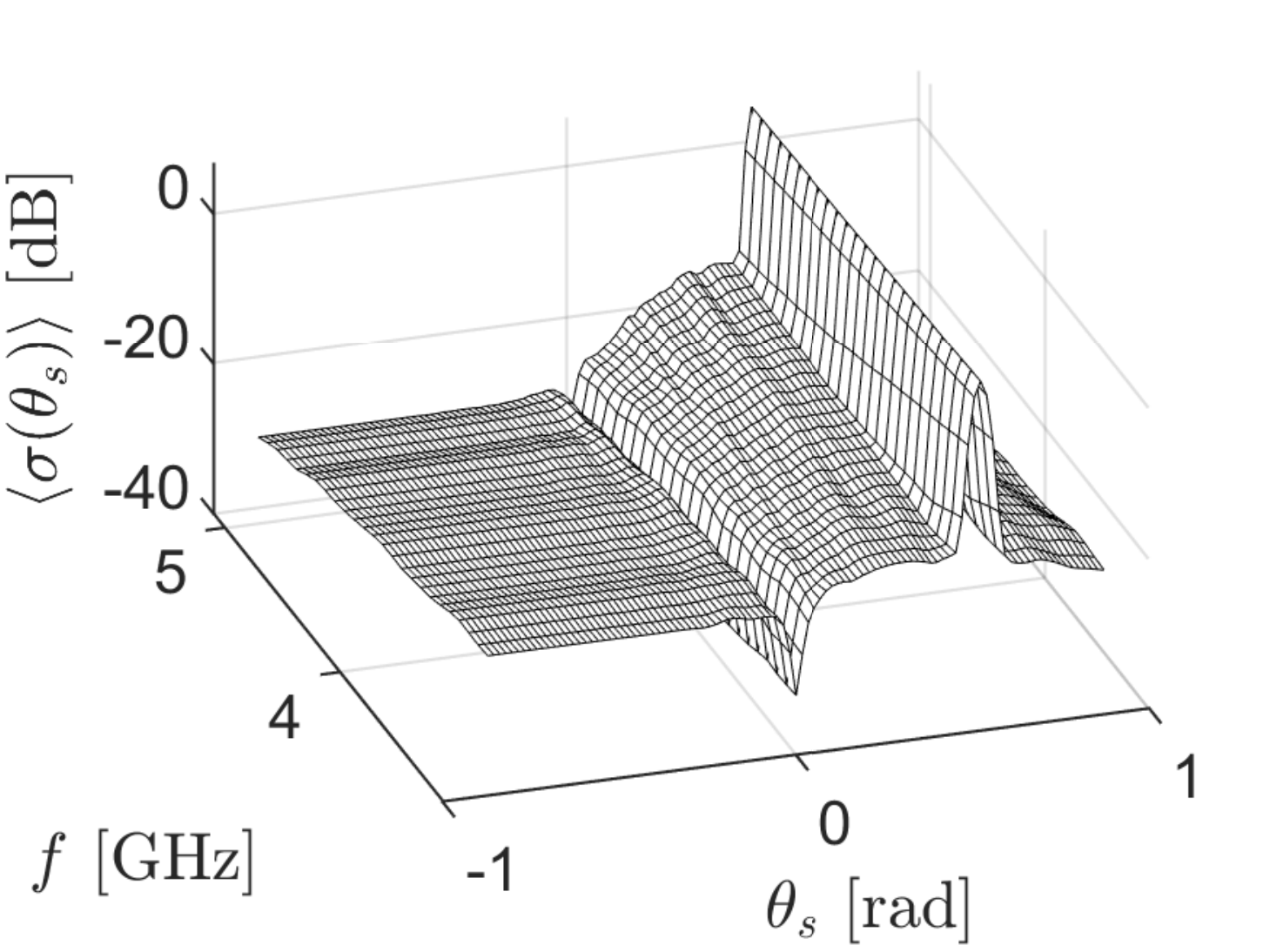}
  \hfill
  \includegraphics[width=0.40\linewidth]{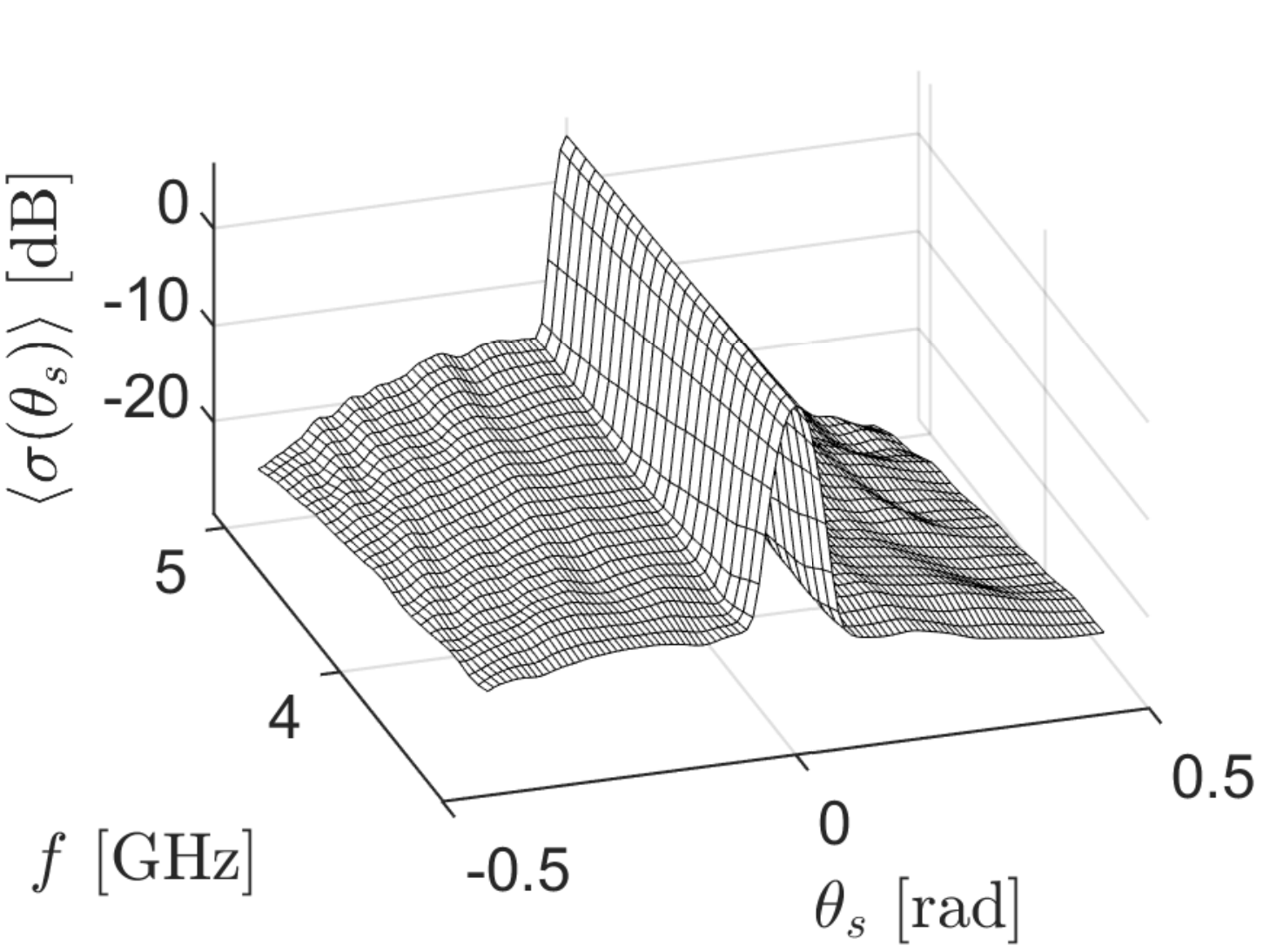}
  \hfill
  \caption{[Left] Average of the bistatic cross-section,
    $\langle \sigma(\theta_{s},\theta_{i}) \rangle$, over $100$
    realizations of a Gaussian-correlated random rough surface with
    $h_{\text{RMS}} = 0.2$ cm and $\ell = 8$ cm due to a plane wave
    incident with $\theta_{i} = 30$ degrees. [Right] A close-up of
    this result about $\theta_{s} = \theta_{i}$.}
  \label{fig:enhanced-backscattering}
\end{figure}

In Fig.~\ref{fig:enhanced-backscattering} we show the bistatic
cross-section due to a plane wave with $\theta_{i} = 30$ degrees
averaged over $100$ realizations of a Gaussian-correlated rough
surface with RMS height $h_{\text{RMS}} = 0.2$ cm and correlation
length $\ell = 8$ cm. These results show a sharp angular cone about
$\theta_{s} = \theta_{i}$ as a consequence of enhanced
backscattering. Enhanced backscattering is a canonical multiple
scattering phenomenon in which counter-propagating scattered waves add
coherently in the retro-reflected direction,
$\theta_{s} = \theta_{i}$.

With these surface roughness parameters, we find that scattering by
the random rough surface is significant and cannot be ignored. Because
these rough surfaces exhibit enhanced backscattering, there is
significant multiple scattering.  Moreover, SAR measurements
use a single emitter/receiver, so we measure the field exactly
at the retro-reflected angle corresponding to the peak of the angular
cone.  However, we do not care to reconstruct this rough surface
profile for this imaging problem. Rather, we seek a method that
attempts to identify and locate targets without needing to consider
this rough surface. Nonetheless, scattering by the rough surface will
be an important factor in the measurements.

\section{Modeling measurements}
\label{sec:measurements}

In this work we consider scattering by subsurface point targets. This
assumption simplifies the modeling of measurements which, in turn,
enables the determination of the effectiveness of a subsurface imaging
method. We consider imaging point targets here as a necessary first
problem for any effective imaging method to solve.

To model measurements we must consider both the ground bounce signal
that is the reflection by the rough surface, and the scattered signal
by the targets. Assuming that scattering by each target is independent
from any others, we give the procedure we use to model measurements
for a single point target located at $(x_{1},z_{1})$ below due to a
point source located at $(x_{0},z_{0})$.

\begin{enumerate}

\item Compute one realization of the Gaussian-correlated rough
  surface, $z = h(x)$, with RMS height $h_{\text{RMS}}$ and correlation
  length $\ell$.

\item Solve the system \eqref{eq:BIE-top}. Let $U_{0}$ and $V_{0}$
  denote the solution.

\item Compute the ground-bounce signal, $R$, through evaluation of
  \begin{equation*}
    R = \mathscr{D}_{0}[U_{0}](x_{0},z_{0}) -
    \mathscr{S}_{0}[V_{0}](x_{0},z_{0}).
  \end{equation*}
  This expression is the field reflected by the rough surface
  evaluated at the same location as the source.

\item Solve the system \eqref{eq:BIE-bottom}. Let $U_{1}$ and $V_{1}$
  denote the solution.

\item Compute the field scattered by the point target, $S$, through
  evaluation of
  \begin{equation*}
    S = \left( \mathscr{D}_{0}[U_{1}](x_{0},z_{0}) -
      \mathscr{S}_{0}[V_{1}](x_{0},z_{0}) \right) \rho \left(
      - \mathscr{D}_{1}[U_{0}](x_{1},z_{1}) +
      \mathscr{S}_{1}[V_{0}](x_{1},z_{1}) \right).
  \end{equation*}
  There are three factors in this expression written in right-to-left
  order just like matrix products. The third factor corresponds to the
  field emitted from the source that transmits across the interface
  and is incident on the target. The second factor is the reflectivity
  of the target $\rho$. The first factor is the propagation of the
  second and third terms from the target location to the receiver
  location.

\end{enumerate}
Steps 2 through 5 of this procedure are repeated over each frequency
$\omega_{m}$ for $m = 1, \dots, M$ and each spatial location of the
platform $\xpos_{n}$ for $n = 1, \dots, N$. The results are
$M \times N$ matrices $R$ and $S$. When there are multiple targets, we
repeat Steps 4 and 5 for each of the targets and $S$ is the sum of
those results.

Using this procedure above, we model measurements according to
\begin{equation}
  D = R + S + \eta,
  \label{eq:data-matrix}
\end{equation}
with $\eta$ denoting additive measurement noise which we model as
Gaussian white noise. The inverse scattering problem is to identify
targets and determine their locations from the data matrix $D$.

\section{Ground bounce signal removal}
\label{sec:ground-bounce}

According to measurement model \eqref{eq:data-matrix}, the ground
bounce signal $R$ is added to the scattered signal $S$. The ground
bounce signal does not contain any information about the
targets. Since we do not seek to reconstruct the interface for this
imaging problem, $R$ impedes the solution of the inverse scattering
problem. Hence, we seek to remove it from measurements.

The key assumption we make is that the relative amount of power in $R$
is larger than that in $S$. This assumption opens the opportunity to
use principal component analysis to attempt to remove $R$ from
$D$. Let $D = U \Sigma V^{H}$ denote the singular value decomposition
of $D$ where $V^{H}$ denotes the Hermitian or conjugate transpose of
$V$. Because of uncertainty in the interface, we are not able to
explicitly determine the structure of the singular values $\sigma_{j}$
for $j = 1, \dots, \min(M,N)$ in the $M \times N$ diagonal matrix
$\Sigma$. Instead we seek to observe any changes in the spectrum of
singular values that indicate a separation between contributions by
$R$ and $S$.

Consider $M = 25$ frequencies uniformly sampling the bandwidth ranging
from $3.1$ GHz to $5.1$ GHz and $N = 21$ spatial locations of the
platform uniformly sampling the aperture $a = 1$ m at $1$ m above the
mean interface height $\langle h(x) \rangle = 0$. We set
$\epsilon_{r} = 9$ and $\beta = 0.1$. Using one realization of a rough
surface with $h_{\text{RMS}} = 0.2$ cm and $\ell = 8$ cm, we compute
$R$. Then we compute the SVD of $R$ and examine the singular values.

\begin{figure}[htb]
  \centering
  \includegraphics[width=0.4\linewidth]{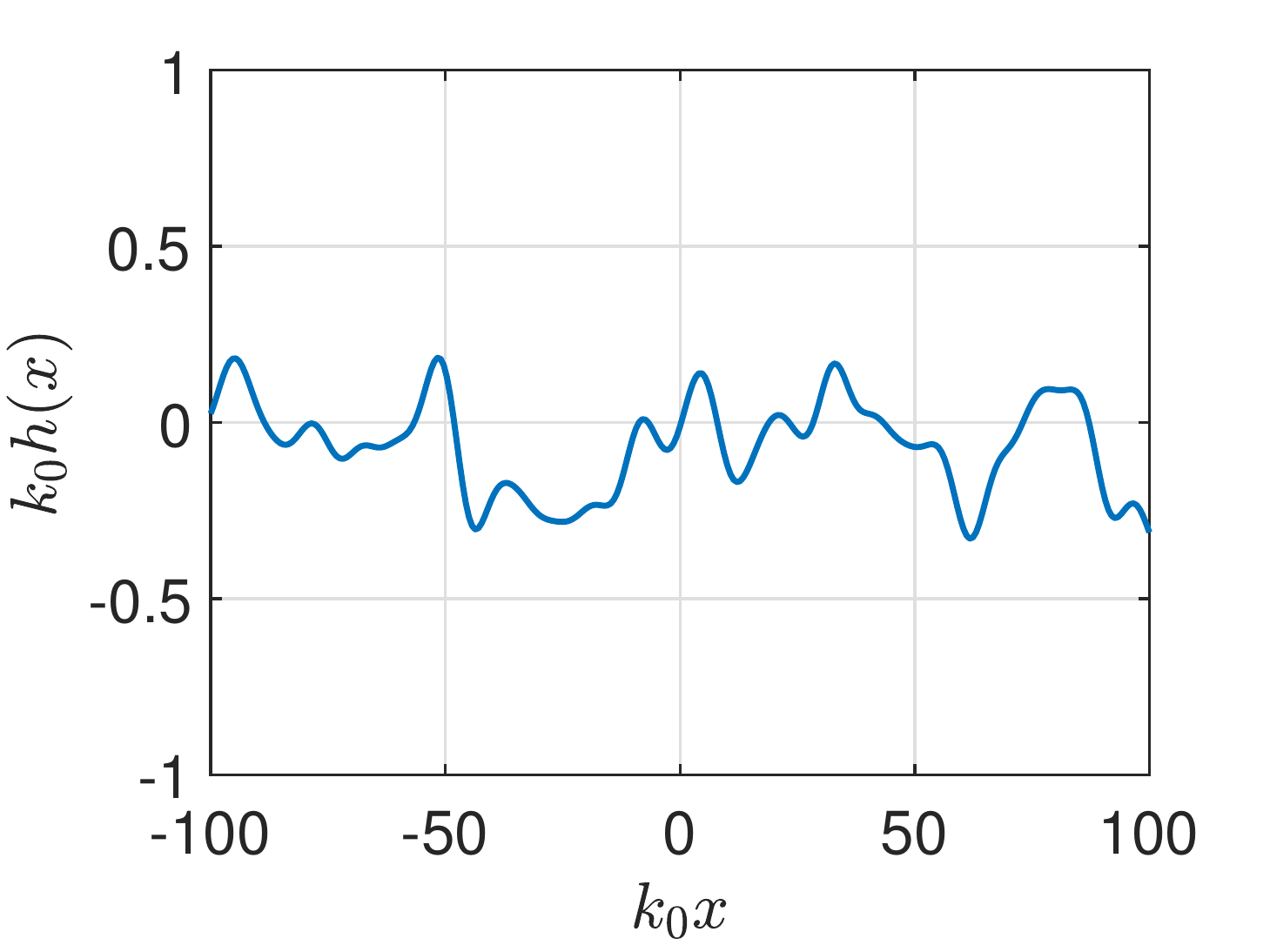}
  \includegraphics[width=0.4\linewidth]{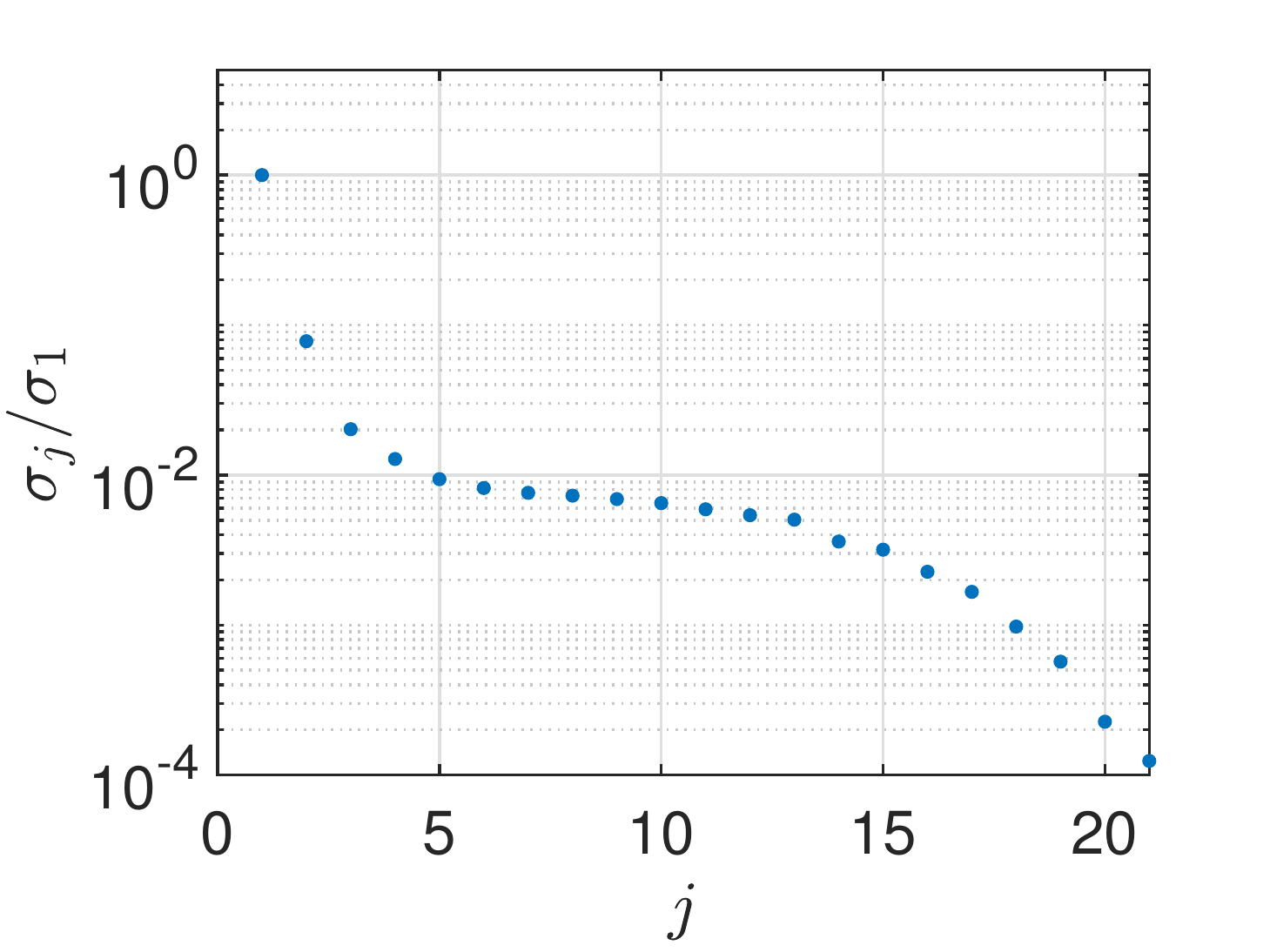}
  \caption{[Left] One realization of the Gaussian-correlated random
    rough surface with $h_{\text{RMS}} = 0.2$ cm and $\ell = 8$ cm
    with $k_{0}$ denoting the wavenumber at the central
    frequency. [Right] The singular values of the ground bounce
    signals by this rough surface normalized by the first singular
    value $\sigma_{1}$.}
  \label{fig:R-SVD}
\end{figure}

In Fig.~\ref{fig:R-SVD} we show results for one realization of the
Gaussian-correlated rough surface with $h_{\text{RMS}} = 0.2$ cm and
$\ell = 8$ cm shown in the left plot and the corresponding singular
values (normalized by the first singular value, $\sigma_{1}$) for the
resulting ground bounce signals in the right plot. Note that this
realization of the rough surface is one among those used to study the
bistatic cross-section in Fig.~\ref{fig:enhanced-backscattering} which
exhibited enhanced backscattering. Consequently, we know that the
ground bounce signals include strong multiple scattering by the rough
surface.

Looking at the singular values in Fig.~\ref{fig:R-SVD} we identify a
change in behavior in their decay. From $j = 1$ to $j = 5$, we find
that $\sigma_{j}$ decays rapidly over two orders of magnitude. In
contrast, from $j = 6$ to $j \approx 15$, we find that the decay of
$\sigma_{j}$ is much slower and then decays thereafter. We have
observed that this qualitative behavior of the singular values
persists over different realizations.

Through these observations of the behavior of singular values for $R$,
we now propose a method to approximately remove $R$ from $D$ given as
the following procedure.
\begin{enumerate}

\item Compute the SVD of the measurement matrix $D = U \Sigma V^{H}$.

\item Identify the index $j^{\ast}$ where the rapid decay of the
  singular values stops and the behavior changes.

\item Compute
  \begin{equation}
    \tilde{D} = D - \sum_{i = 1}^{j^{\ast}} \sigma_{i} \mathbf{u}_{i}
    \mathbf{v}_{i}^{H}, 
    \label{eq:D-tilde}
  \end{equation}
  where $\mathbf{u}_{i}$ and $\mathbf{v}_{i}$ denote the $i$-th
  columns of $U$ and $V$, respectively.
  
\end{enumerate}
It is likely that this procedure does not remove $R$ from $D$
exactly. However, we apply this procedure to obtain $\tilde{D}$ and
test below if this procedure works well enough for identifying and
locating targets.

Note that measurement noise is applied to $D = R + S$. The
corresponding SNR is defined according to
$\text{SNR} = 10 \log_{10}( \| R + S \|_{F} / \| \eta \|_{F} )$ with
$\| \cdot \|_{F}$ denoting the Frobenius norm. This SNR is dominated
by $R$ since $\| R \|_{F} \gg \| S \|_{F}$. When we remove $R$ from
$D$, there will be an effective SNR
($\text{eSNR} = 10 \log_{10} ( \| S \|_{F}^{2} / \| \eta \|_{F}^{2}
)$) based on $S$ which will be much lower. For this reason, we see
that this subsurface imaging problem is more sensitive to noise than
other imaging problems where ground bounce signals are not present.

\section{Kirchhoff migration imaging}
\label{sec:KM}

Consider a sub-region of $z < h(x)$ where we seek to form an image.
We call this sub-region the imaging window (IW). Let
$(x,z) \in \mathrm{IW}$ denote a search point in the IW. To form an
image which identifies targets and gives estimates for their
locations, we evaluate the KM imaging functional,
\begin{equation}
  I^{\text{KM}}(\ypos) = \left| \sum_{m = 1}^{M} \sum_{n = 1}^{N}
    \tilde{d}_{mn} a_{mn}^{\ast}(x,z) \right|,
  \label{eq:KM}
\end{equation}
over a mesh of grid points sampling the IW. Here $\tilde{d}_{mn}$ is
the $(m,n)$ entry of the matrix $\tilde{D}$ and $a_{mn}(x,z)$ are
called the illuminations. The superscript $^{\ast}$ denotes the
complex conjugate. The illuminations effectively back-propagate the
data so that the resulting image formed shows peaks on the target
locations.

\subsection{Computing illuminations}

To compute the illuminations $a_{mn}(x,z)$ we first note that we do
not know the interface $z = h(x)$ nor do we seek to reconstruct
it. However, we assume that $\langle h(x) \rangle = 0$ is known, so we
consider the interface $z = 0$ instead. Additionally, we do not know
the loss tangent $\beta$ that dictates the absorption in the lower
medium. In fact, we have shown previously that making use of any
knowledge of the absorption is not useful for imaging to identify and
locate targets~\cite{KT-Lossy}. However, we assume that
$\epsilon_{r}$ is known. With these assumptions, we write
\begin{equation}
  a_{mn}(x,z) = \phi^{(0)}_{mn}(x,z) \phi^{(1)}_{mn}(x,z).
\end{equation}
Here, $\phi^{(0)}_{mn}(x,z)$ corresponds to the field on $(x,z)$ due
to a point source with frequency $\omega_{m}$ located at $\xpos_{n}$
whose amplitude is normalized to unity. The quantity
$\phi^{(1)}_{mn}(x,z)$ is the field with frequency $\omega_{m}$
evaluated on $\xpos_{n}$ due to a point source at $(x,z)$ whose
amplitude is normalized to unity.

Using Fourier transform methods, we find that the field $u^{(0)}$
evaluated on $(x,z)$ due to a point source with frequency $\omega_{m}$
located at $\xpos_{n} = (x_{n},z_{n})$ is
\begin{equation}
  u^{(0)} = \frac{\mathrm{i}}{2\pi} \int 
    \frac{e^{\mathrm{i} ( q_{0} z_{n} - q_{1} z ) }}{q_{0} + q_{1}} 
  e^{\mathrm{i} \xi ( x - x_{n} )} \mathrm{d}\xi,
\end{equation}
with $q_{0} = \sqrt{\omega_{m}^{2}/c^{2} - \xi^{2} }$ and
$q_{1} = \sqrt{\epsilon_{r} \omega_{m}^{2} /c^{2} -
  \xi^{2} }$. Similarly, we find that the field $u^{(1)}$ evaluated on
$(x_{n}, z_{n})$ due to a ponit source with frequency $\omega_{m}$
located at $(x,z)$ is
\begin{equation}
  u^{(1)} = \frac{\mathrm{i}}{2 \pi} \int \frac{e^{\mathrm{i} ( q_{0}
      z_{n} - q_{1} z)}}{q_{0} + q_{1}} e^{\mathrm{i} \xi ( x_{n} - x
    )} \mathrm{d}\xi.
\end{equation}
Upon computing $u^{(0)}$ and $u^{(1)}$, we evaluate
$\phi_{mn}^{(0)} = u^{(0)}/| u^{(0)} |$ and
$\phi_{mn}^{(1)} = u^{(1)}/| u^{(1)} |$.

Both $u^{(0)}$ and $u^{(1)}$ are integrals of the form,
\begin{equation}
  I = \int_{-\infty}^{\infty} \frac{f(\xi)}{\sqrt{k_{0}^{2} - \xi^{2}}
    + \sqrt{ k_{1}^{2} - \xi^{2}}} e^{\mathrm{i} \beta_{1}
    \sqrt{k_{0}^{2} - \xi^{2}} + \mathrm{i} \beta_{2} \sqrt{ k_{1}^{2}
      - \xi^{2} }} e^{\mathrm{i} \xi \gamma} \mathrm{d}\xi,
  \label{eq:FourierIntegral}
\end{equation}
with $k_{1} = k_{0} \sqrt{\varepsilon_{r}}$, and $\beta_{1}$,
$\beta_{2}$, and $\gamma$ denoting real parameters.  The wavenumbers
$k_{0}$ and $k_{1}$ are real, and we assume that $|k_{0}| < |k_{1}|$.
This Fourier integral, which is one example of a Sommerfeld integral,
is notoriously difficult to compute due to the highly oscillatory
behavior of the function inside the integral. There have been several
approaches to compute this Fourier integral accurately
\cite{cai2002algorithmic, o2014efficient, bruno2016windowed}. To
compute (\ref{eq:FourierIntegral}), we follow \cite{barnett2011new}
and integrate on a deformed contour in the complex plane to avoid
branch points.  Here, we use the deformed contour
\begin{equation*}
  \xi(s) = s + \mathrm{i} A \left[ e^{-w (s + k_{0})^{2}} + e^{-w (s +
      k_{1})^{2}} - e^{-w (s - k_{0})^{2}} - e^{-w (s -
      k_{1})^{2}} \right],
\end{equation*}
with $-\infty < s < \infty$, and $A$ and $w$ denoting user-defined
parameters. Integration is taken with respect to $s$ over a truncated,
finite interval chosen so that the truncation error is smaller than
the finite precision arithmetic. In the simulations that follow, we
have used $500$ quadrature points with $A = 0.4$ and $w = 6$.  We also
use the suggestion in \cite{barnett2011new} of applying the mapping
$s = \sinh( \beta )$ with $-\infty < \beta < \infty$ to cluster
quadrature points in the interval $(-k_{0}, k_{0})$.

\subsection{Modified KM}

We have recently developed a modification to KM that allows for
tunably high-resolution images of individual
targets~\cite{KimTsogka-Tunable}. Suppose that we have evaluated
\eqref{eq:KM} and identified a target. In a region about that target,
we normalize $I^{\mathrm{KM}}$ so that its peak value is $1$. Let
$\bar{I}^{\mathrm{KM}}$ denote the normalization of $I^{\mathrm{KM}}$
in this region. With this normalized image, we compute the following
M\"obius transformation,
\begin{equation}
  I_{\delta}^{\text{KM}}(\ypos) = \frac{\delta}{ 1 - ( 1 -
    \delta ) \bar{I}^{\mathrm{KM}}(\ypos)},
  \label{eq:rKM}
\end{equation}
with $\delta > 0$ denoting a user-defined tuning parameter. We call
the resulting image formed with (\ref{eq:rKM}) the modified KM
image. In the whole space, we have determined that this modified KM
method scales the resolution of KM by $\sqrt{\delta}$. Because
$\delta$ is a user-defined quantity, it can be set to be arbitrarily
small. It is in this way that $I_{\delta}^{\text{KM}}$ produces
tunably high-resolution images of targets.

\section{Numerical results}
\label{sec:numerics}
We now present numerical results where we have (i) simulated
measurements using the procedure given in Section
\ref{sec:measurements}, (ii) removed the ground bounce signal using
the procedure given in Section \ref{sec:ground-bounce}, and then
produced images through evaluation of the KM and modified KM imaging
functions given in Section \ref{sec:KM}.

Just as we have done for the results shown in Section
\ref{sec:ground-bounce}, we have used $M = 25$ frequencies uniformly
sampling the bandwidth ranging from $3.1$ GHz to $5.1$ GHz and
$N = 21$ spatial locations of the platform uniformly sampling the
aperture $a = 1$ m situated $1$ m above the average interface height
$\langle h(x) \rangle = 0$. We set $\epsilon_{r} = 9$ and
$\beta = 0.1$ as suggested by Daniels for modeling buried
landmines~\cite{daniels:2006}. We compute imaging results for one
realization of a Gaussian-correlated rough surface that has
$h_{\text{RMS}} = 0.2$ cm and $\ell = 8$ cm.

\begin{figure}[t]
  \centering
  \includegraphics[width=0.4\linewidth]{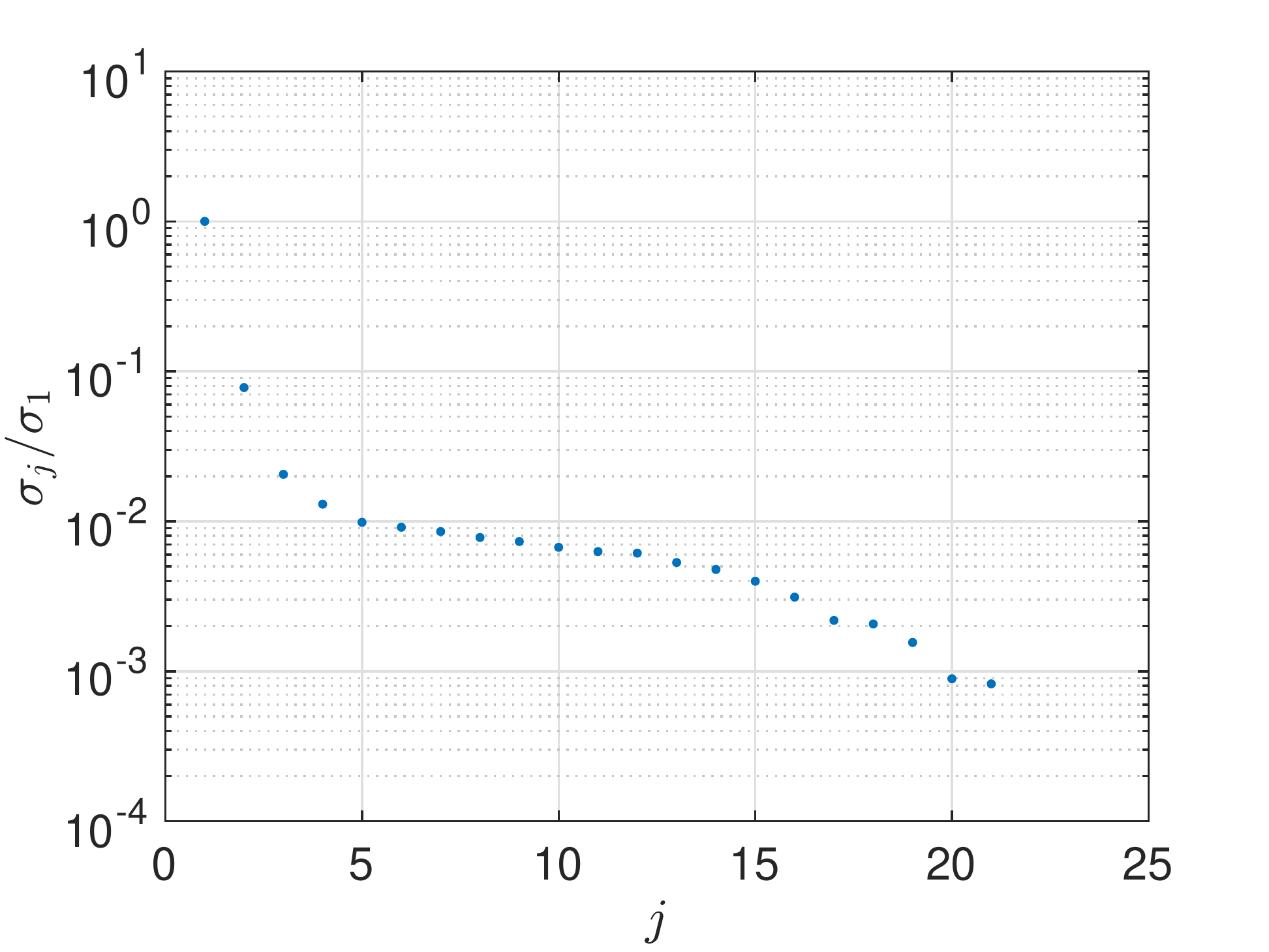}
  \caption{Singular values of the matrix $D$. These measurements
    include the ground bounce signals by one realization of a
    Gaussian-correlated rough surface with $h_{\text{RMS}} = 0.2$ cm
    and $\ell = 8$ cm. Additionally, they include scattering by a point
    target located at $(2,-8)$ cm with $\rho = 3.4
    \mathrm{i}$. Measurement noise has been added so that
    $\text{SNR} = 24.2$ dB.}
  \label{fig:singular-values}
\end{figure}

\subsection{Single target}

Let the origin of a coordinate system correspond to the center of the
flight path in the $x$-coordinate and the mean surface height
$\langle h(x) \rangle = 0$ in the $z$-coordinate as shown in
Fig.~\ref{fig:SAR-problem}. We compute images for a target located at
$(2,-8)$ cm with reflectivity $\rho = 3.4 \mathrm{i}$. Measurement
noise is added to the simulated measurements so that
$\text{SNR} = 24.2$ dB.

\begin{figure}[t]
  \centering
  \begin{subfigure}[t]{0.32\linewidth}
    \includegraphics[width=\linewidth]{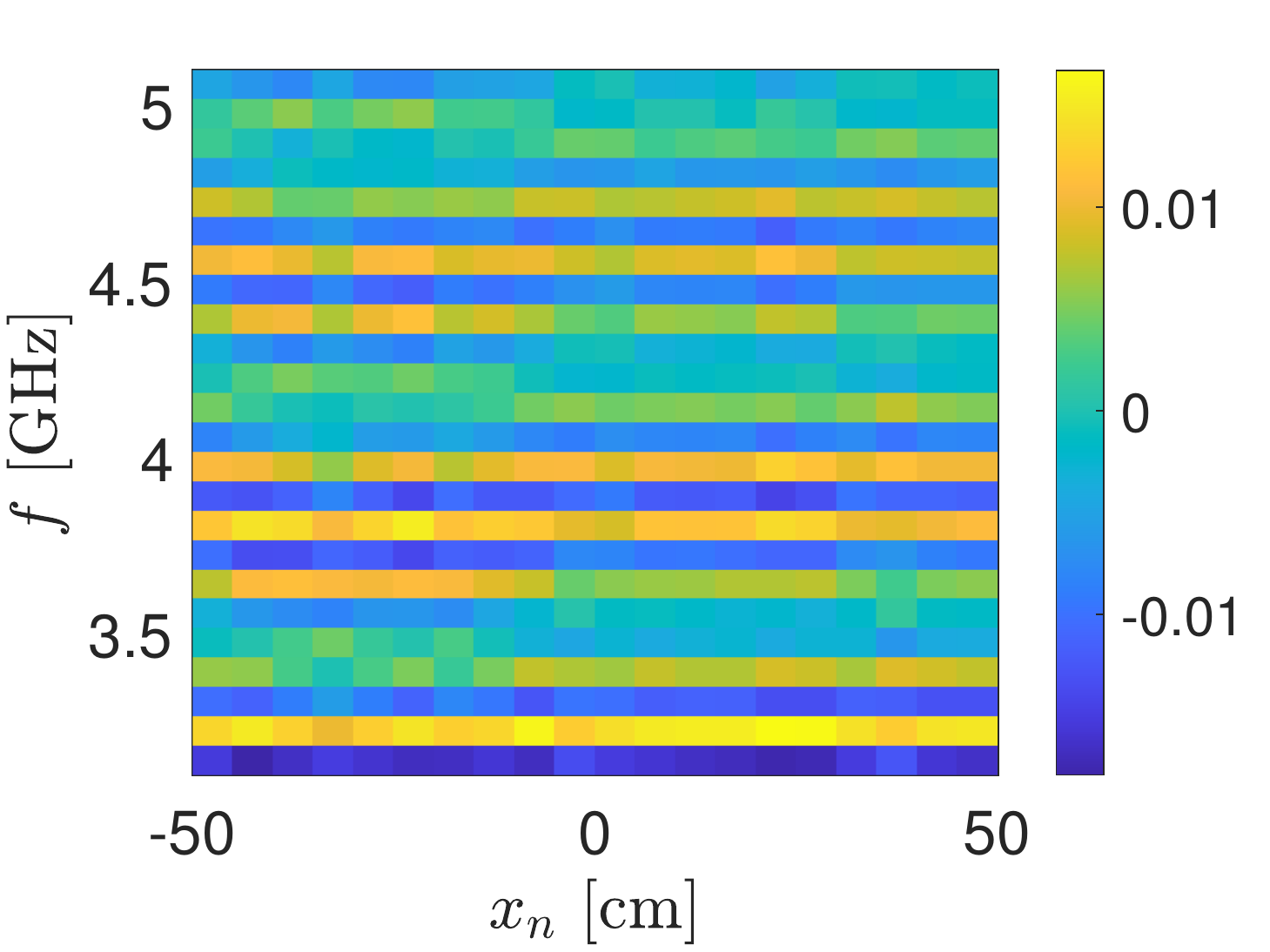}
    \caption{$D$}
  \end{subfigure}
  \hspace{0.5cm}
  \begin{subfigure}[t]{0.32\linewidth}
    \includegraphics[width=\linewidth]{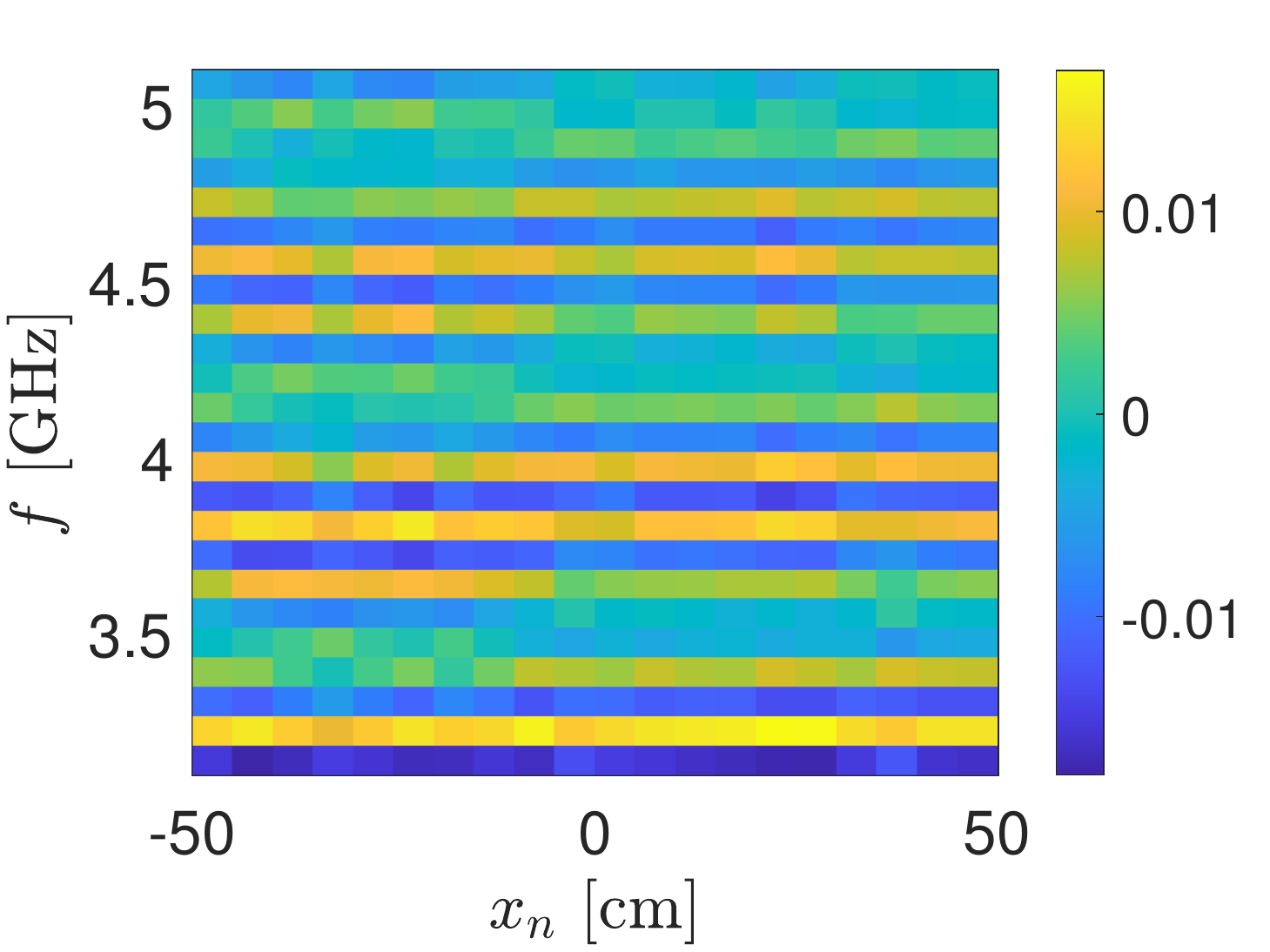}
    \caption{$R$}
  \end{subfigure}
  \\
  \begin{subfigure}[t]{0.32\linewidth}
    \includegraphics[width=\linewidth]{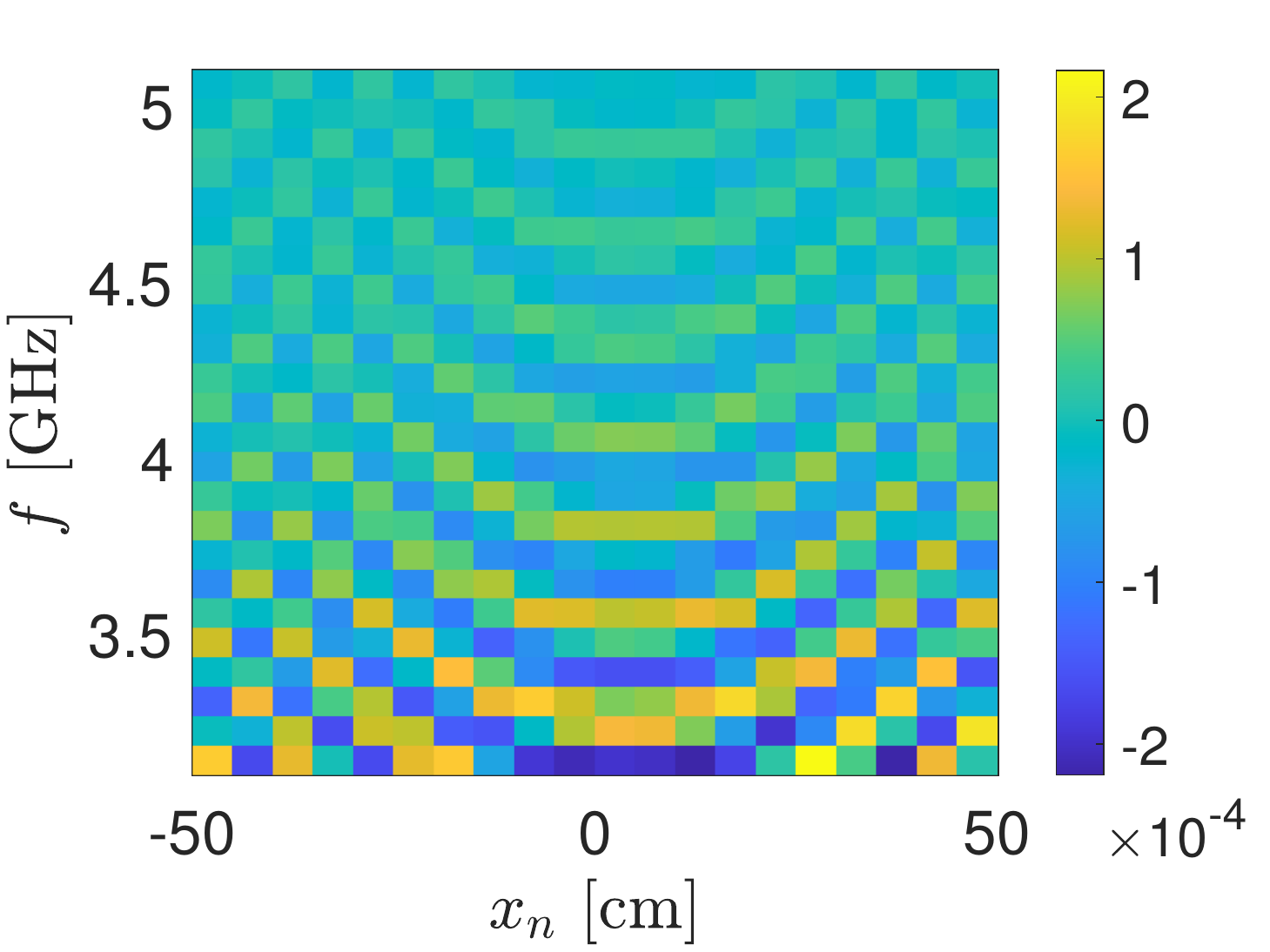}
    \caption{$S$}
  \end{subfigure}
  \hspace{0.5cm}
  \begin{subfigure}[t]{0.32\linewidth}
    \includegraphics[width=\linewidth]{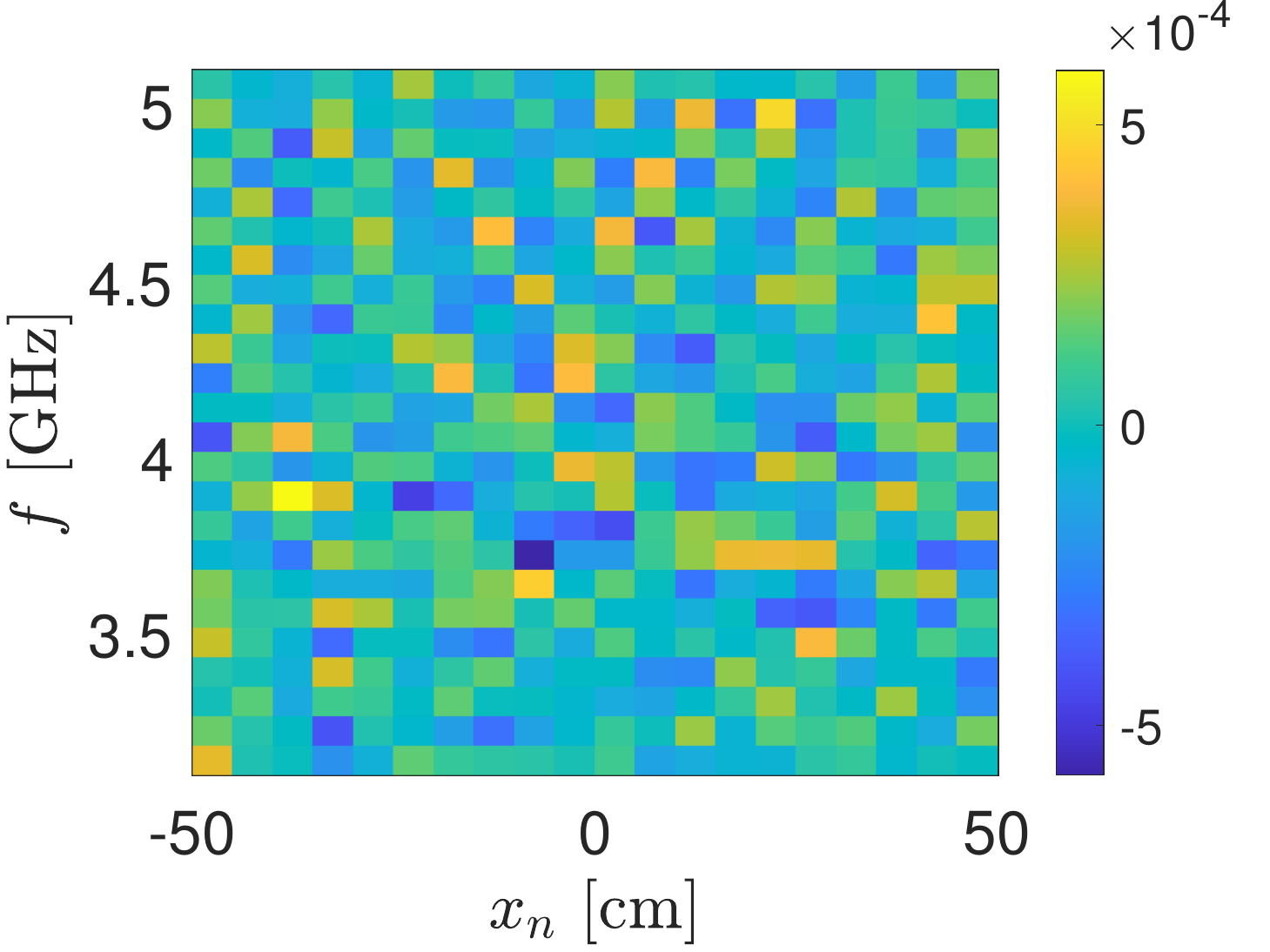}
    \caption{$\tilde{D}$}
  \end{subfigure}
  \caption{Real part of the entries of (a) the data matrix $D$, (b)
    the ground bounce signals $R$, (c) the scattered signals $S$, and
    (d) the matrix $\tilde{D}$ with the contributions from the first
    $5$ singular values removed.}
  \label{fig:data-matrices}
\end{figure}

Figure \ref{fig:singular-values} shows the singular values for the
data matrix $D$ normalized by the first singular value. Similar to
what we observed in Section \ref{sec:ground-bounce} with the ground
bounce signals, we find that the first $5$ singular values decay
rapidly. The singular values $\sigma_{j}$ for $j > 5$ show a different
behavior. Thus, we apply the ground bounce removal procedure given in
Section \ref{sec:ground-bounce} using $j^{\ast} = 5$. 

We show real part of the data matrix $D$ in the top left plot of
Fig.~\ref{fig:data-matrices}. In the top right plot of
Fig.~\ref{fig:data-matrices} we show the real part of the ground
bounce signals in $R$. Note that the plots for $D$ and $R$ are nearly
indistinguishable consistent with our assumption that the ground
bounce signals dominate the measurements. In the bottom left plot of
Fig.~\ref{fig:data-matrices} we show the real part of the scattered
fields in $S$. Note that those values in $S$ are nearly $2$ orders of
magnitude smaller than those of $R$. The bottom right plot shows the
real part of $\tilde{D}$ resulting from removing the contributions
from the first $j^{\ast} = 5$ singular values. While the magnitudes of
the values in $S$ and $\tilde{D}$ are comparable, they appear
qualitatively different from one another. Thus, it is unclear from
these results whether or not $\tilde{D}$ contains information
regarding the target.

\begin{figure}[t]
  \centering
  \hfill
  \includegraphics[width=0.3\linewidth]{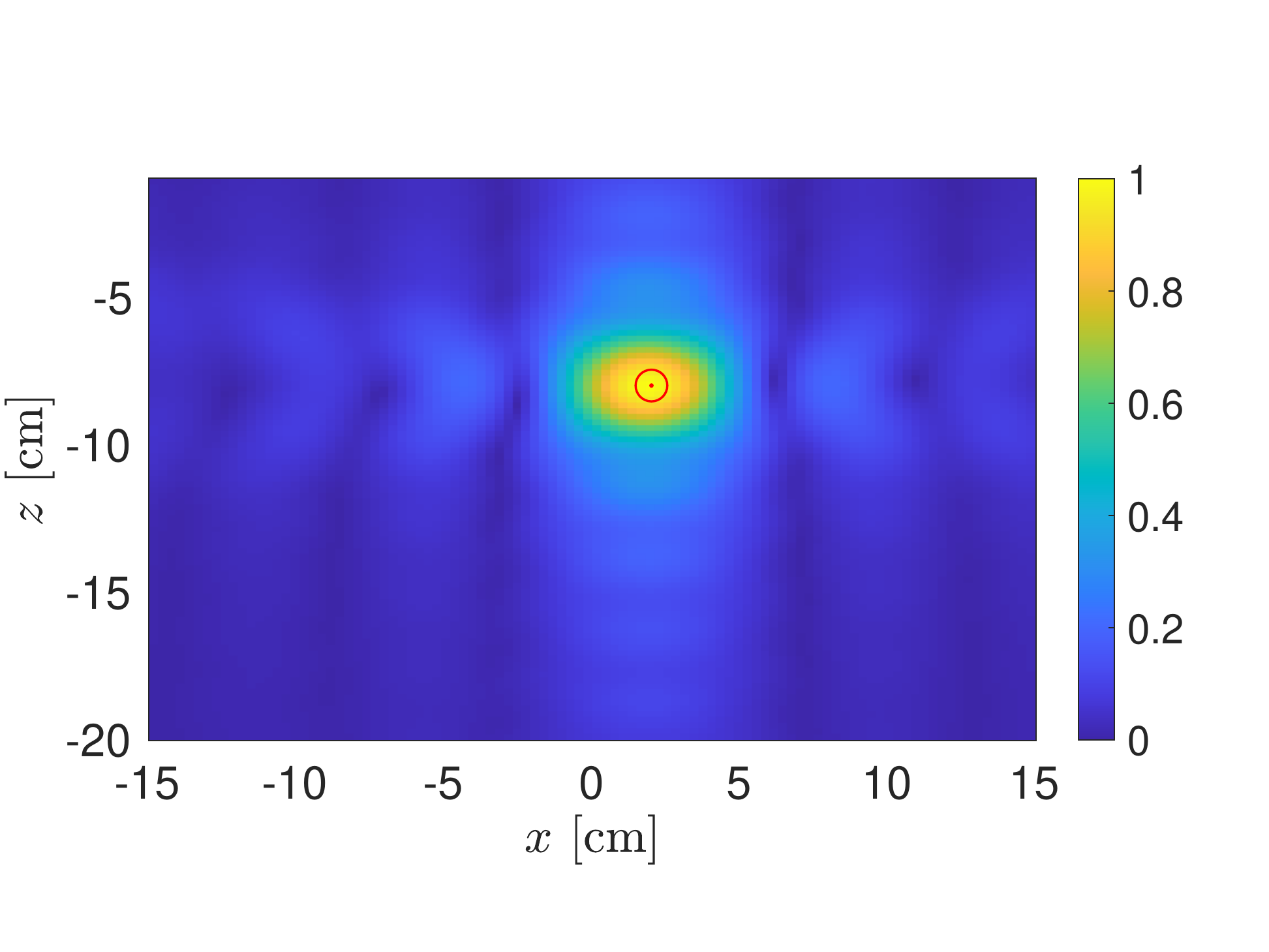}
  \hfill
  \includegraphics[width=0.3\linewidth]{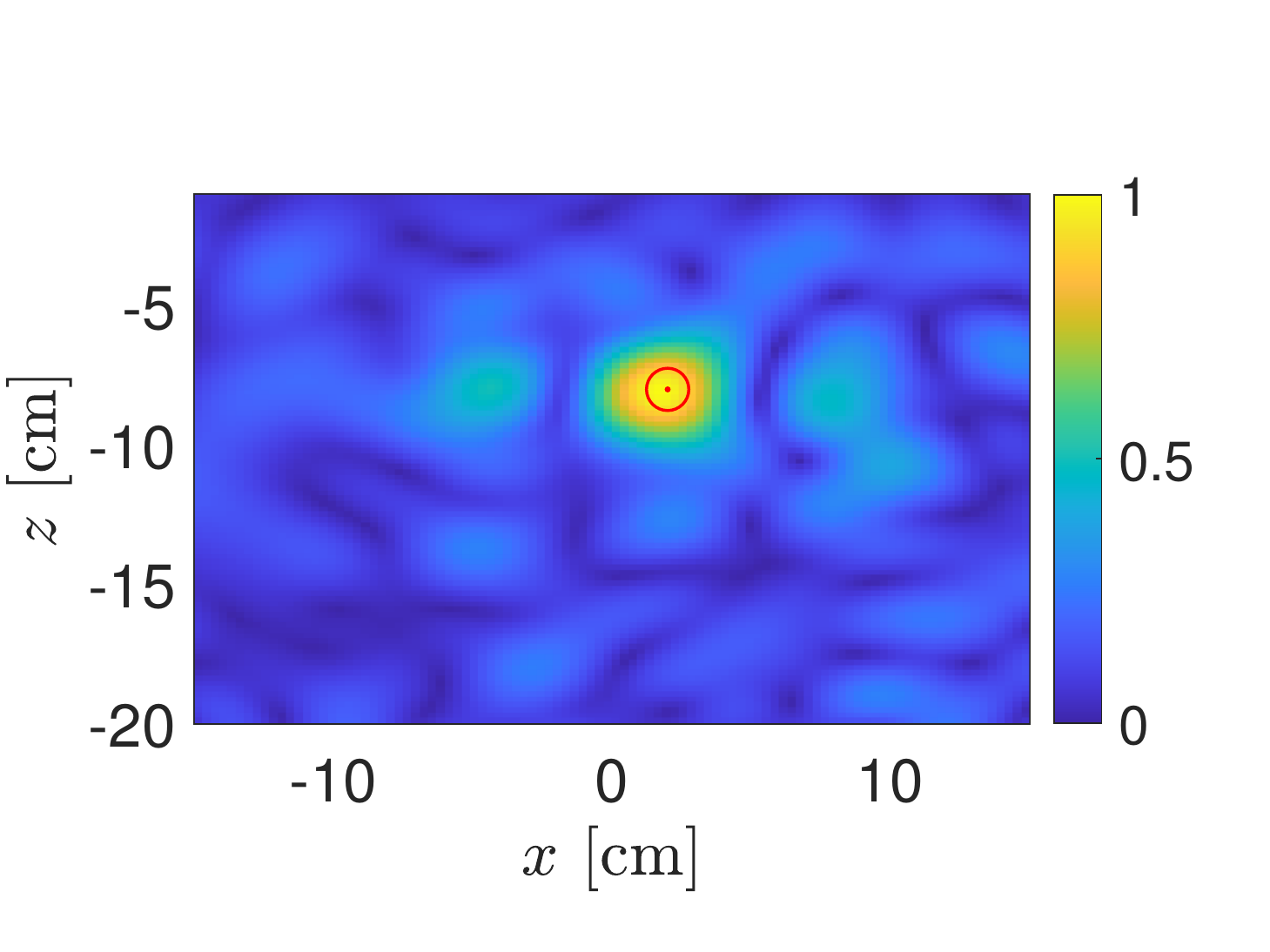}
  \hfill
  \includegraphics[width=0.3\linewidth]{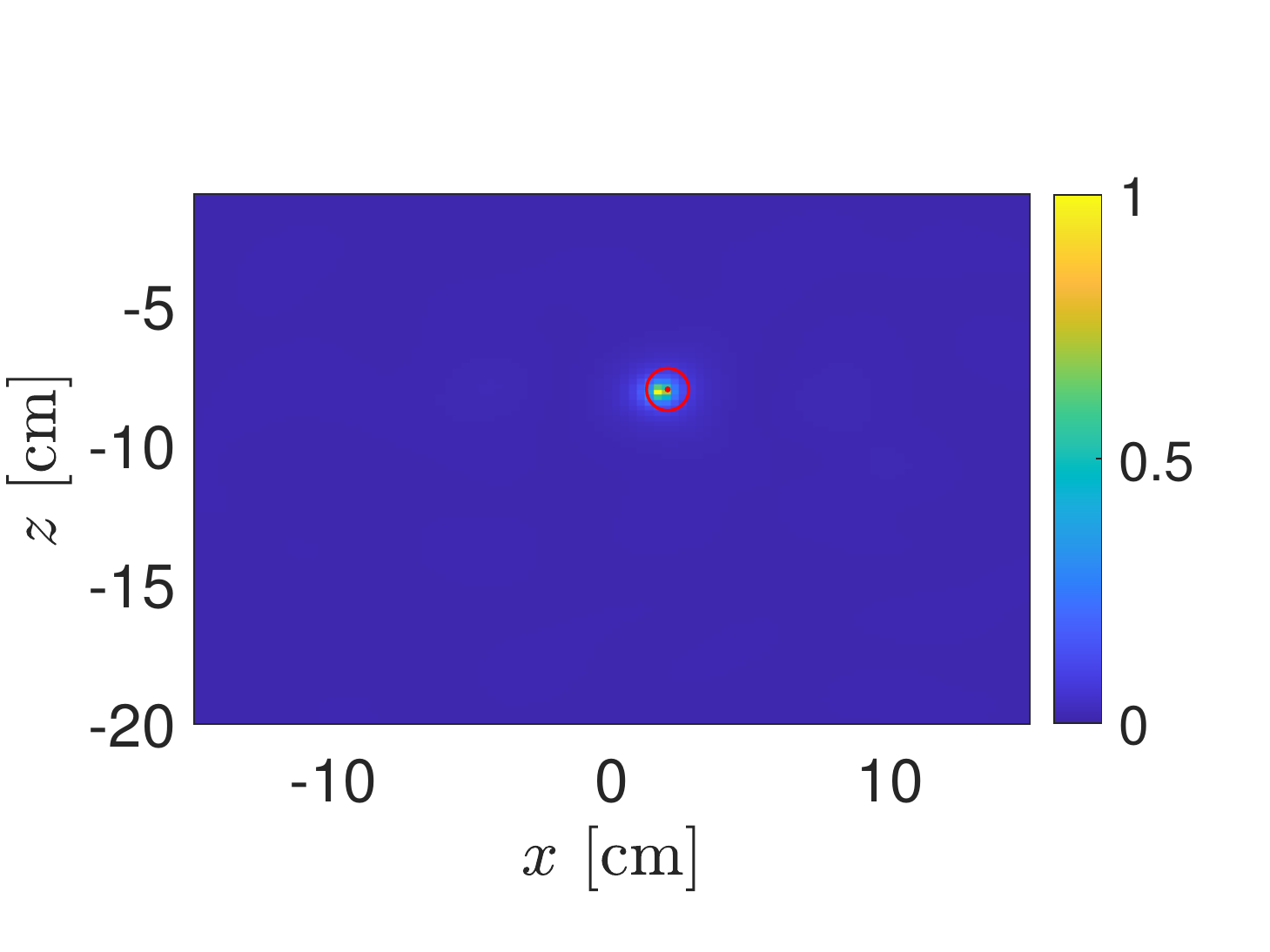}
  \hfill
  \caption{[Left] The ideal imaged formed through evaluation of the KM
    imaging function \eqref{eq:KM} applied to the scattered signals
    contained in $S$. [Center] The image formed through evaluation of
    \eqref{eq:KM} applied to $\tilde{D}$. [Right] The imaged formed
    through evaluation of the modified KM imaging function
    \eqref{eq:rKM} with $\delta = 10^{-2}$ applied to the KM image in
    the center. In each of the plots, the exact target location is
    plotted as a red ``$\odot$'' symbol. }
  \label{fig:images-8cm-depth}
\end{figure}

In Fig.~\ref{fig:images-8cm-depth} we apply KM (center plot) and the
modified KM with $\delta = 10^{-2}$ (right plot) to $\tilde{D}$. For
reference, we have also included the result of applying KM to $S$ in
the left plot of Fig.~\ref{fig:images-8cm-depth}. This ideal case
represents exact ground bounce removal. Despite the fact that the
results for $S$ and $\tilde{D}$ in Fig.~\ref{fig:data-matrices} were
not qualitatively similar, the corresponding KM images in
Fig.~\ref{fig:images-8cm-depth} are quite similar in the vicinity of
the target and show peaks about the target location, $(2,-8)$cm.
The peak of the KM image (center) is accompanied by several imaging
artifacts away from the target location. In contrast, by applying the
modified KM method we eliminate those artifacts and obtain a high
resolution image of the target. We note that the predicted location
determined from where the KM and modified KM images attain their peak
value on the meshed used to plot them is $(1.5,-8.2)$ cm, which is
slightly shifted from the true location. Nonetheless, this result is
quite good given the uncertainty in the surface, the inexact method
for ground bounce removal, unknown absorption, and substantial
measurement noise in the system.

\begin{figure}[htb]
  \centering
  \begin{subfigure}[t]{\linewidth}
    \centering
    \includegraphics[width=0.4\linewidth]{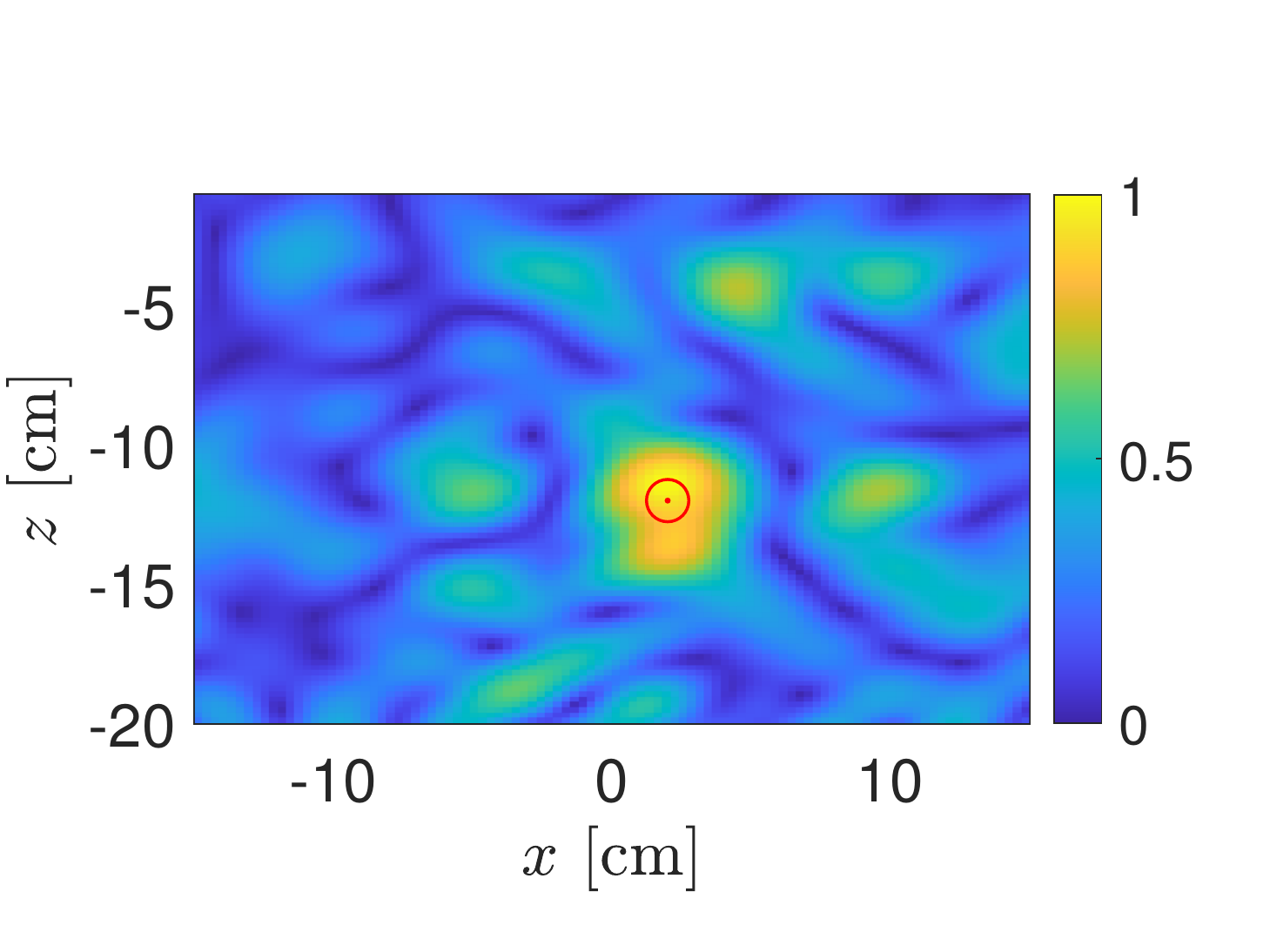}
    \includegraphics[width=0.4\linewidth]{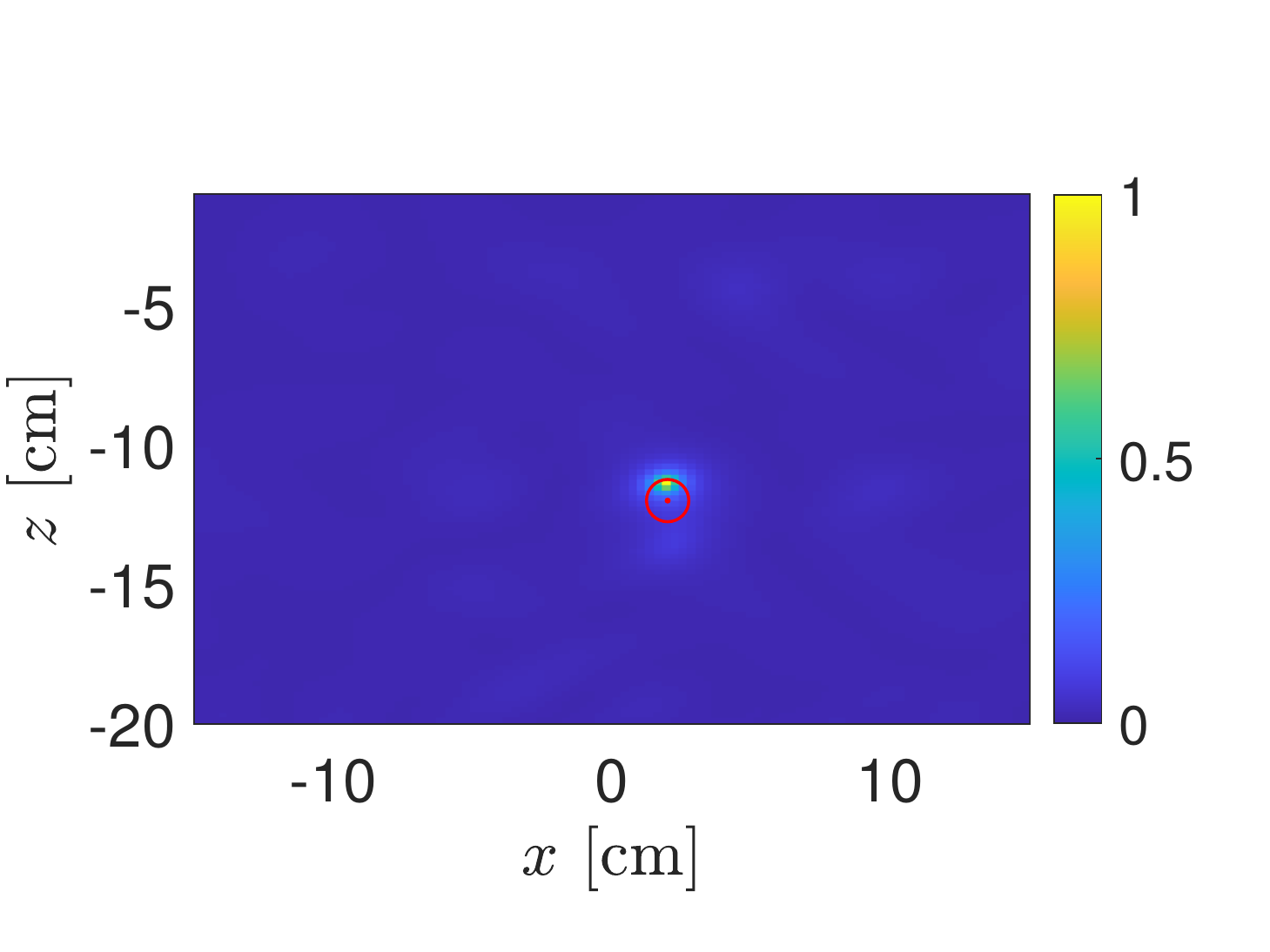}
    \caption{Target located at $(x,z) = (2,-12)$ cm.}
  \end{subfigure}
  \begin{subfigure}[t]{\linewidth}
    \centering
    \includegraphics[width=0.4\linewidth]{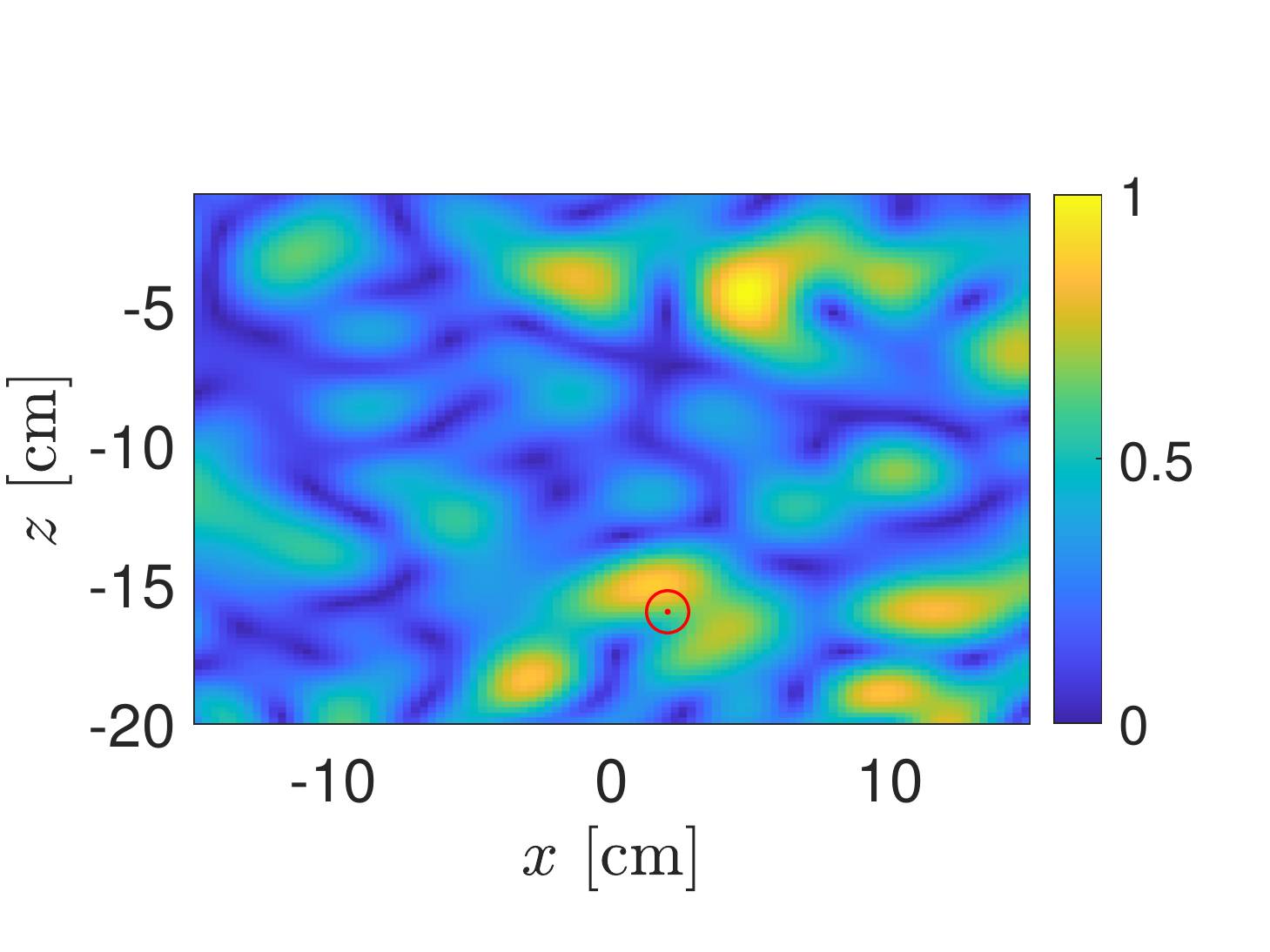}
    \includegraphics[width=0.4\linewidth]{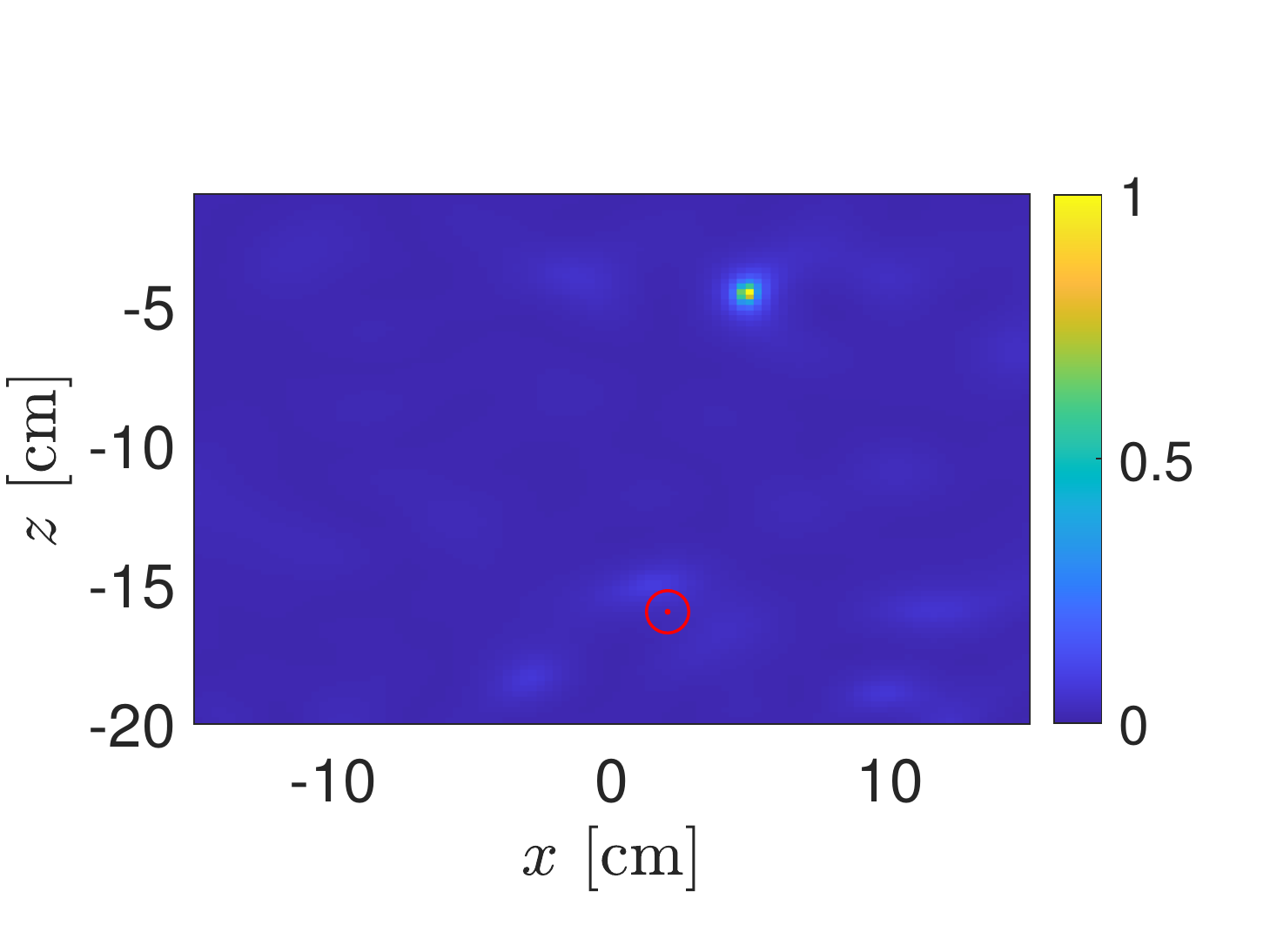}
    \caption{Target at $(x,z) = (2,-16)$ cm.}
  \end{subfigure}
  \caption{[Left] The imaged formed through evaluation of the KM
    imaging function \eqref{eq:KM}. The exact target location is
    plotted as a red ``$\odot$'' symbol. [Right] The imaged formed
    through evaluation of the modified KM imaging function
    \eqref{eq:rKM} with $\delta = 10^{-2}$. The top row is for a
    target located at $(2,-12)$ cm and the bottom row is for a target
    located at $(2,-16)$ cm.}
  \label{fig:depth-study}
\end{figure}

The unknown absorption puts a depth limitation on imaging
targets. When the target depth is comparable to the absorption length,
the imaging method is not able to distinguish between the true target
and a weaker target less deep in the medium. We have observed this
phenomenon with optical diffusion~\cite{GKMT-2018}. Here, uncertainty
in the rough surface complicates this situation even further. In
Fig.~\ref{fig:depth-study} we show KM and modified KM
($\delta = 10^{-2}$) images for a target located at $(2,-12)$ cm (top
row) and for a target located at $(2,-16)$ cm. As the target is placed
deeper into the medium, we observe an increase in the KM imaging
artifacts. For the target located $12$ cm below the surface, we find
that these imaging artifacts contain the peak value of the function
and the target is no longer identifiable in the image. The modified KM
images clearly show this behavior.


\begin{figure}[htb]
  \centering
  \includegraphics[width=0.4\linewidth]{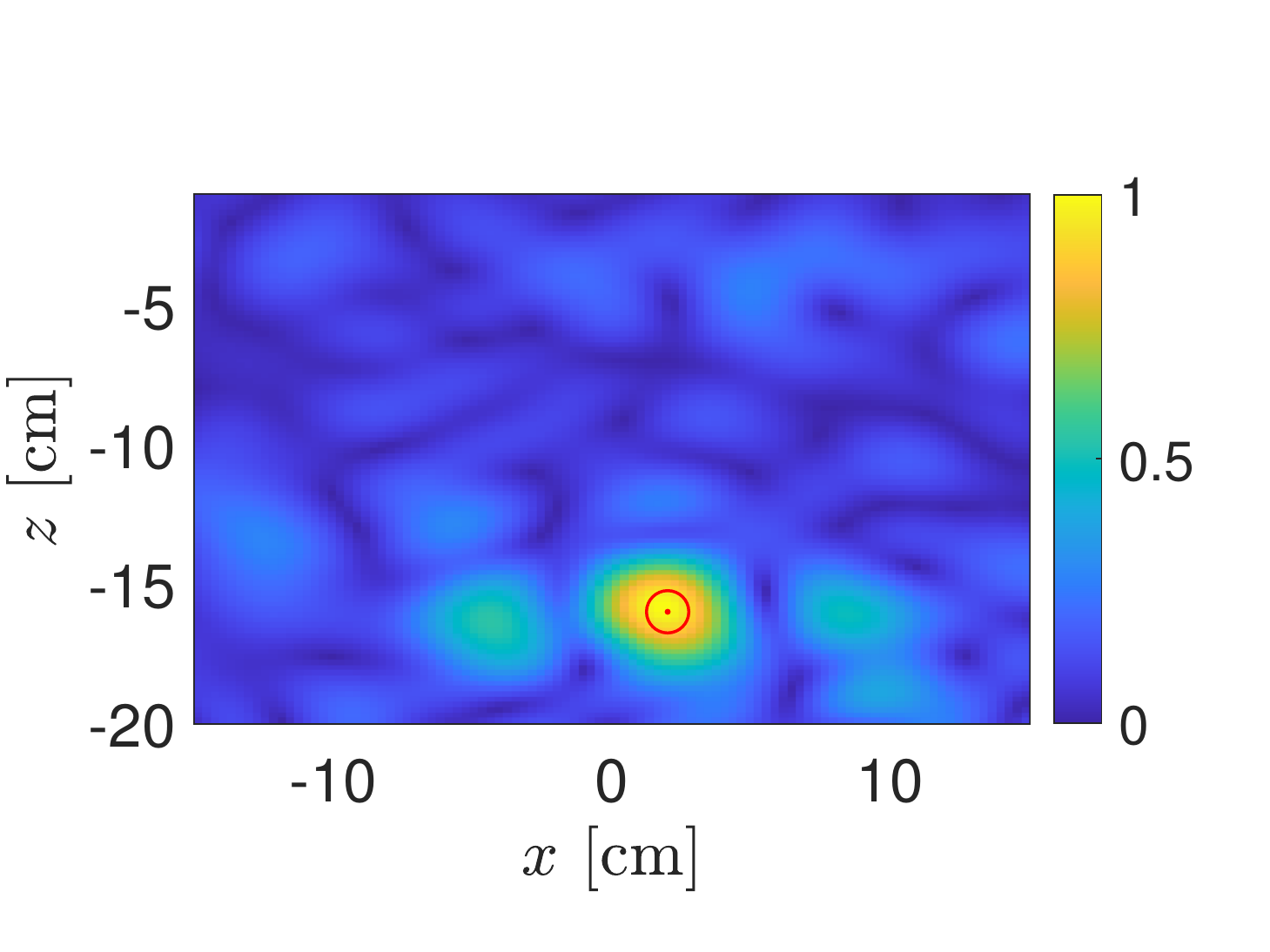}
  \includegraphics[width=0.4\linewidth]{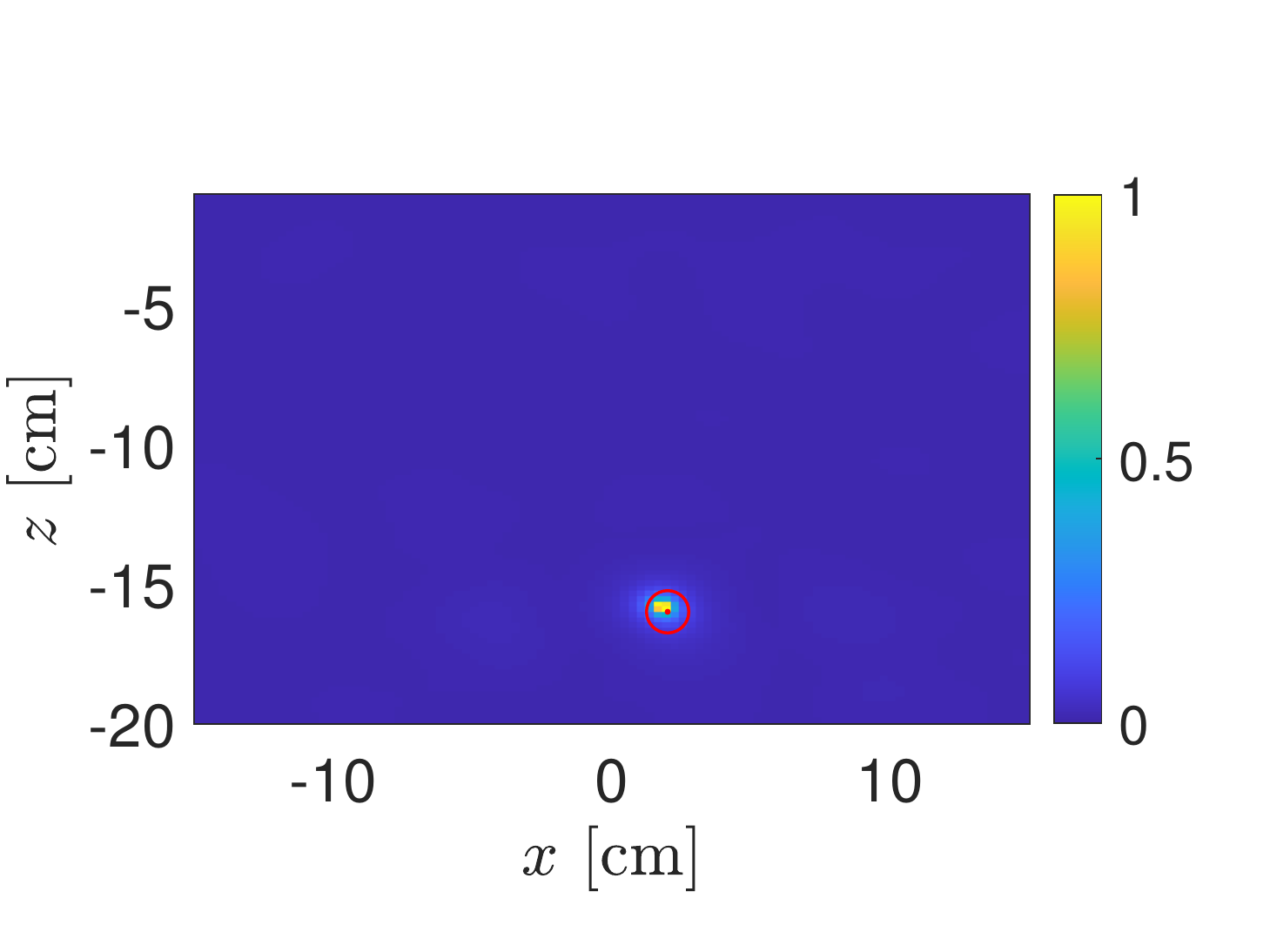}
  \caption{The same as Fig.~\ref{fig:depth-study}(b) except that the
    absorption is reduced from the previous results with
    $\beta = 0.05$.}
  \label{fig:absorption-study}
\end{figure}

The inability of the imaging method to identify targets deep in the
medium is either due to the absorption, the uncertainty of the rough
surface, some combination of these, or possibly other factors. In
Fig.~\ref{fig:absorption-study} we show the resulting image for a
target located at $(2,-16)$ cm with the reduced loss tangent,
$\beta = 0.05$. All other parameters are the same as those used in the
previous images. With this reduced loss tangent, we find that KM and
the modified KM are clearly able to identify the target. From this
result we conclude that the absorption is the main factor limiting the
range of target depths for this imaging method.

\begin{figure}[htb]
  \centering
  \includegraphics[width=0.4\linewidth]{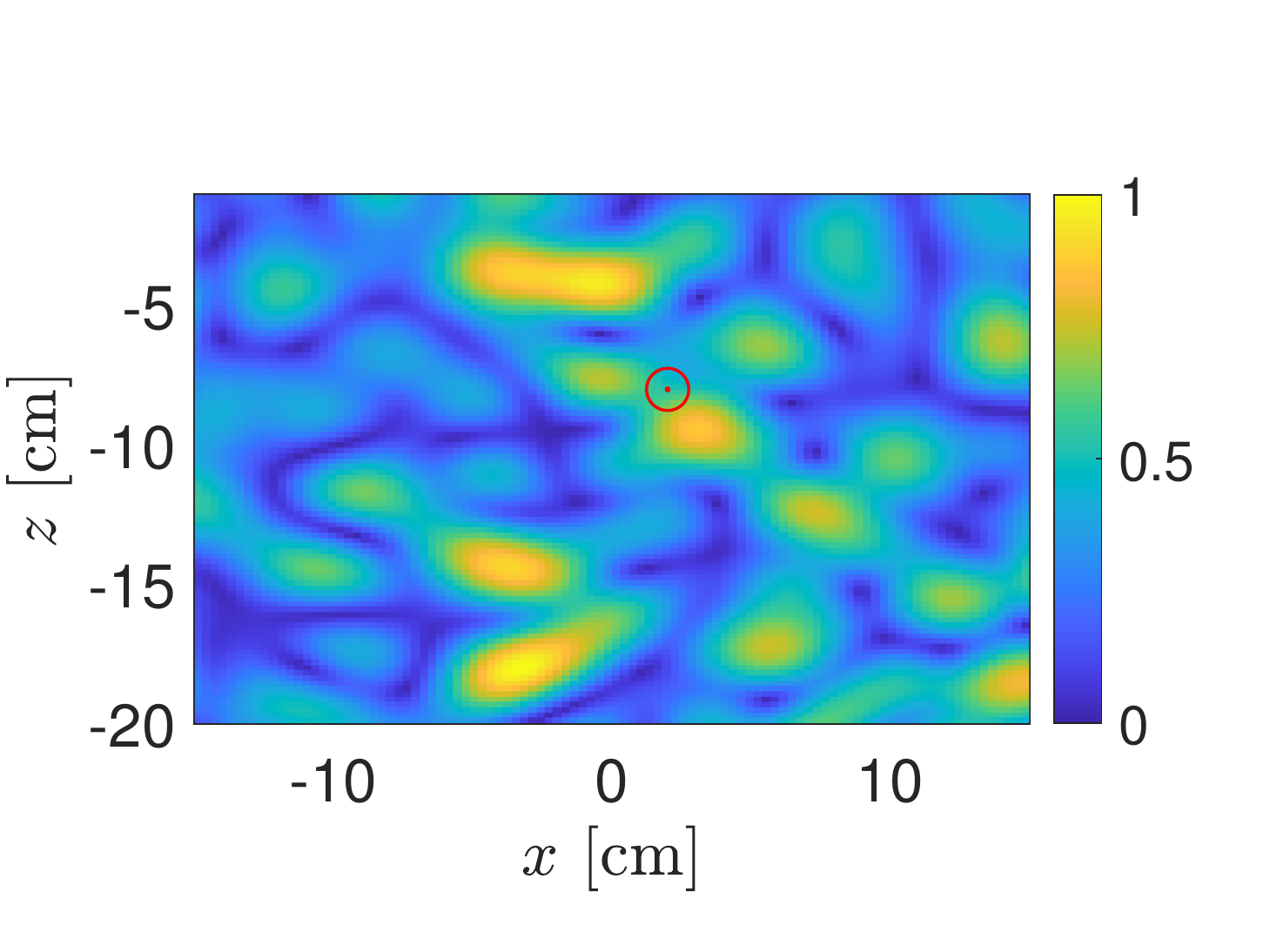}
  \includegraphics[width=0.4\linewidth]{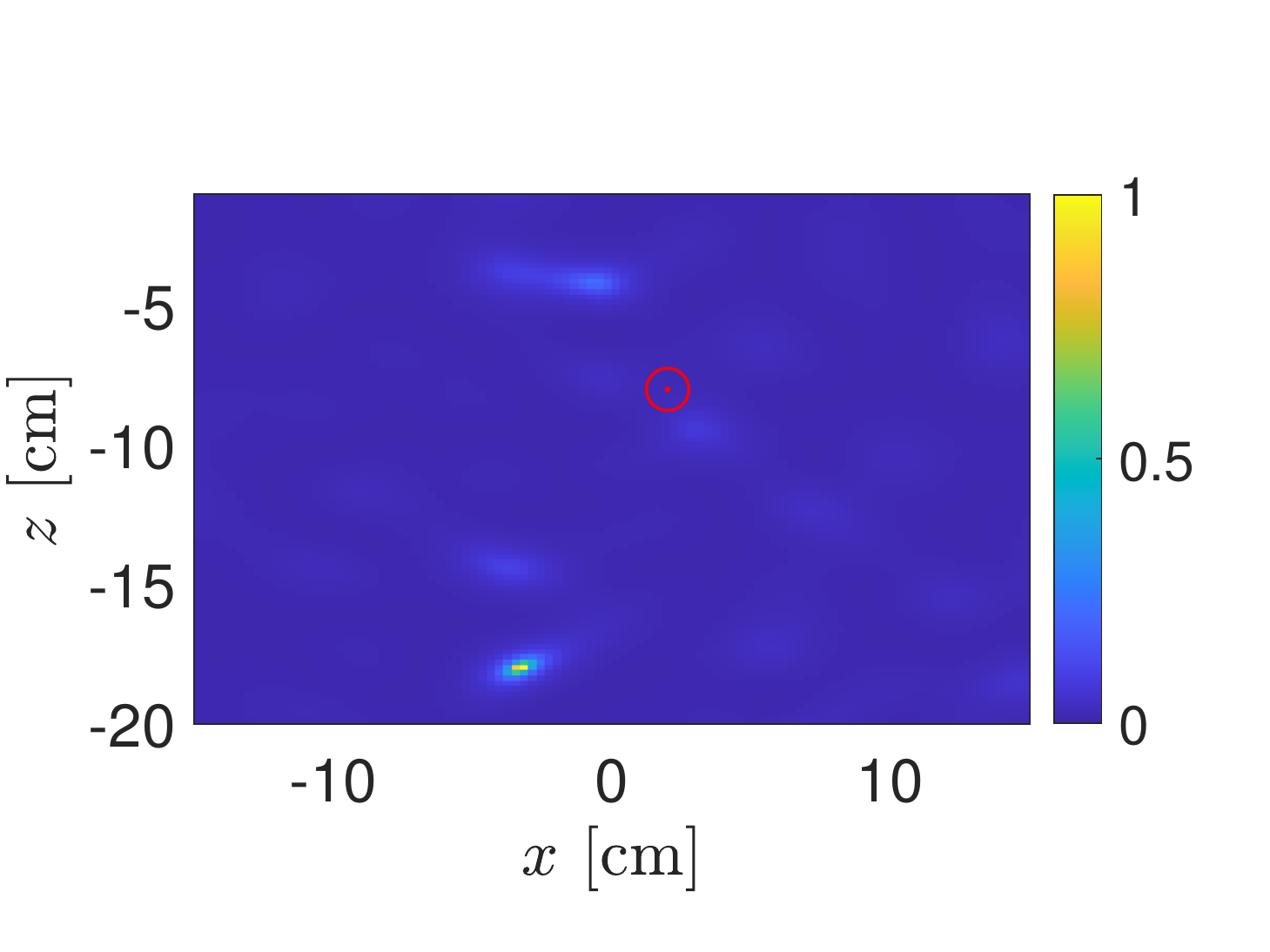}
  \caption{[Left] KM image and [Right] modified KM image with
    $\delta = 10^{-2}$ for a target located at $(2,-8)$ cm with
    $\text{SNR} = 14.2$ dB and $\text{eSNR} = -7.0$ dB.}
  \label{fig:eSNR}
\end{figure}

As we explained above, when we remove ground bounce signals, we
introduce an effective SNR (eSNR) that is important for subsurface
imaging. We expect that KM will be effective as long as
$\text{eSNR} > 0$ dB.  For the results shown in
Fig.~\ref{fig:images-8cm-depth}, $\text{SNR} = 24.2$ dB and
$\text{eSNR} = 3.0$ dB. The resulting image clearly identifies
the target and accurately predicts its location. In contrast, we show
results for $\text{SNR} = 14.2$ dB and $\text{eSNR} = -7.0$ dB in
Fig.~\ref{fig:eSNR}. This image has several artifacts that dominate
over any peak formation about the target location. It is important to
note that the eSNR that we use here cannot be estimated {\it a
  priori}. This result demonstrates that SNR demands on imaging
systems are higher for subsurface imaging problems than other imaging
problems that do not involve ground bounce signals.

\subsection{Multiple targets}

We now consider imaging regions with $3$ targets. Target $1$ is
located at $(-9.0,10.1)$ cm with reflectivity
$\rho_{1} = 3.6 \mathrm{i}$, target $2$ is located at $(1.0,-9.4)$ cm
with reflectivity $\rho_{2} = 3.4 \mathrm{i}$ and target $3$ is
located at $(11.0,-9.8)$ cm with reflectivity
$\rho_{3} = 3.6 \mathrm{i}$. The measurements were computed using the
procedure given in Section \ref{sec:measurements}. Measurement noise
has been added so that $\text{SNR} = 24.2$ dB.

\begin{figure}[htb]
  \centering
  \includegraphics[width=0.4\linewidth]{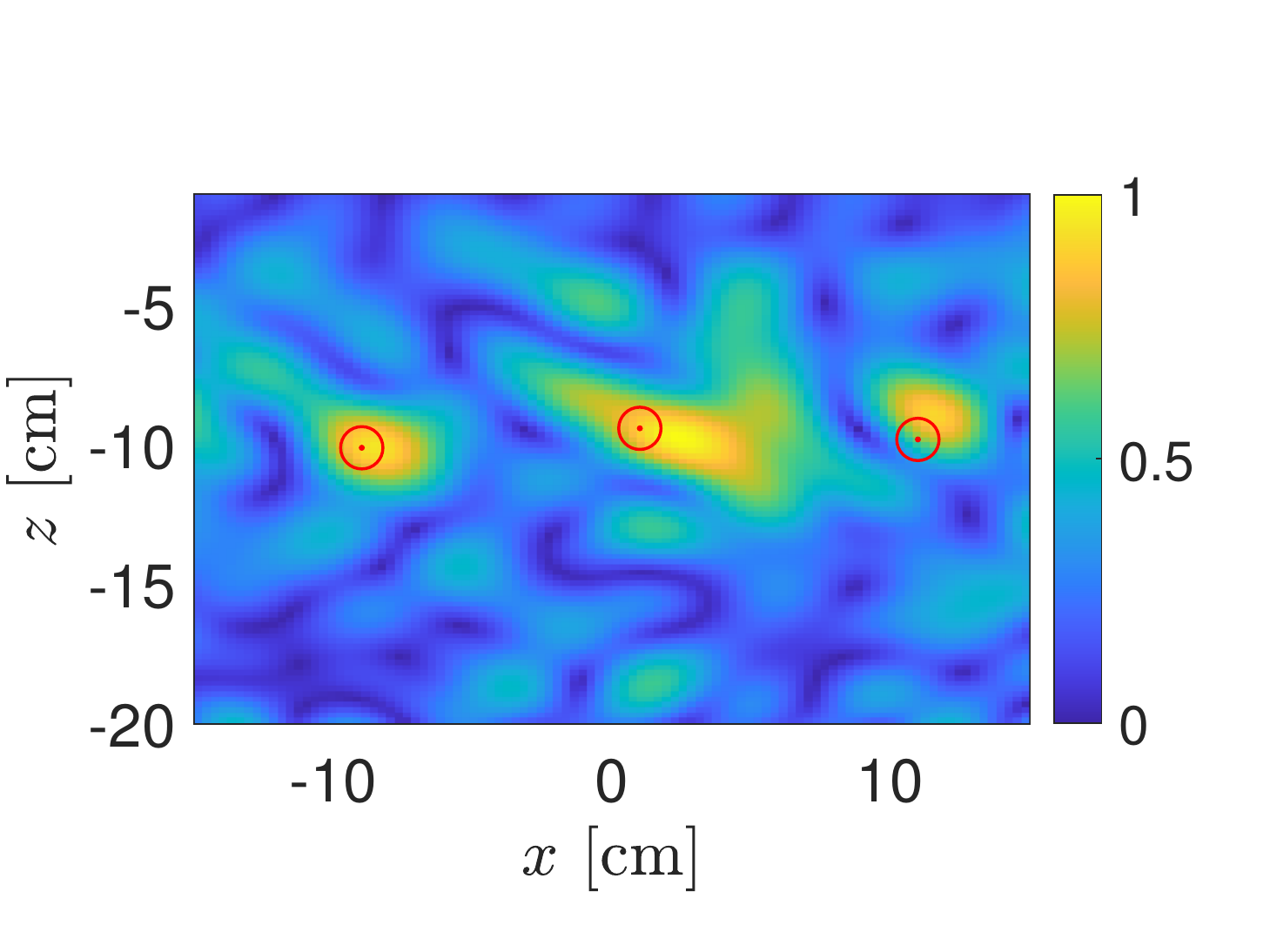}
  \includegraphics[width=0.4\linewidth]{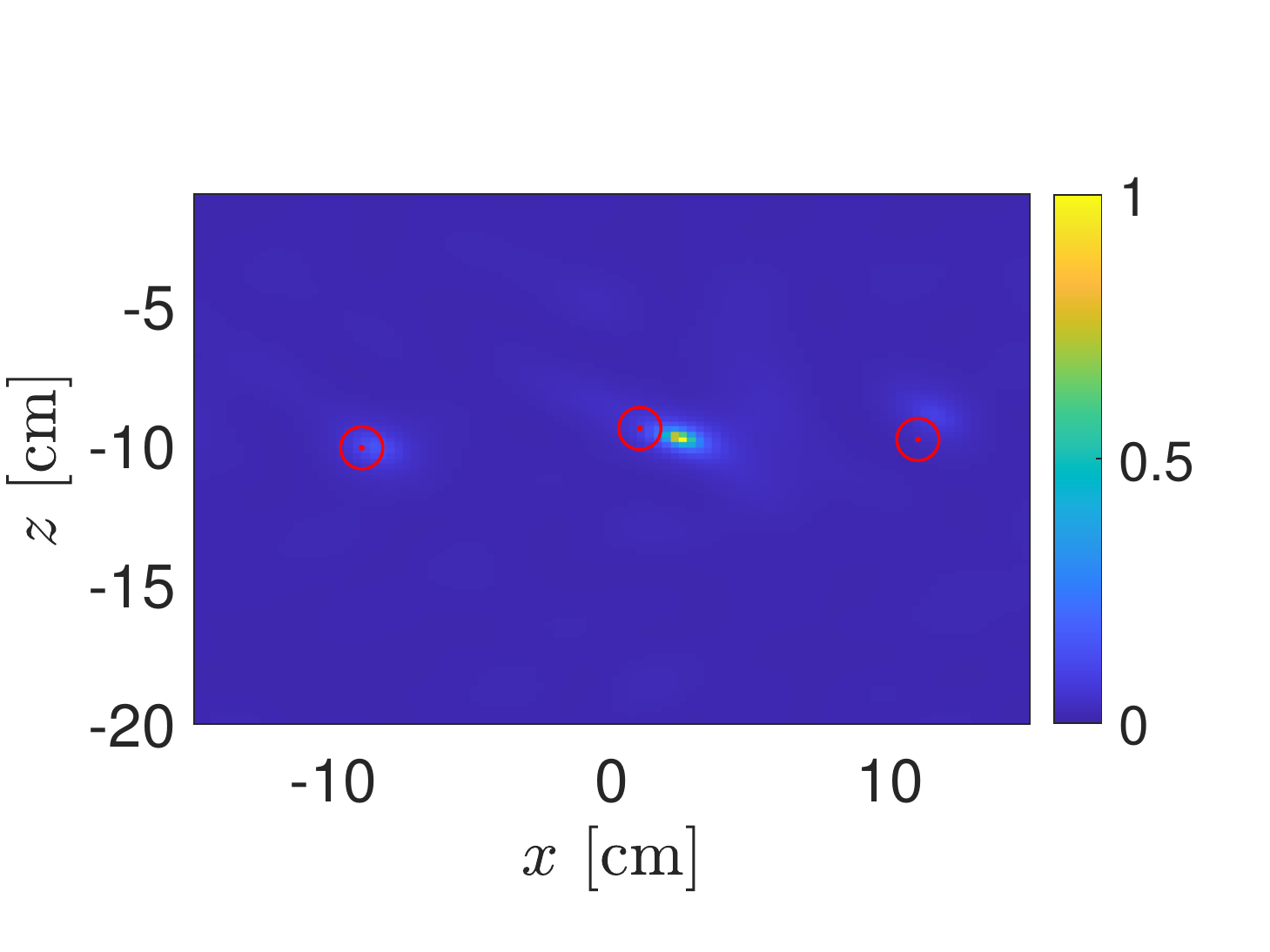}
  \caption{[Left] The imaged formed through evaluation of the KM
    imaging function \eqref{eq:KM} for three targets. The exact target
    locations are plotted as a red ``$\odot$'' symbol. [Right] The image
    formed through evaluation of the modified KM imaging function
    \eqref{eq:rKM} with $\delta = 10^{-2}$. Measurement noise is added
    so that $\text{SNR} = 24.2$ dB.}
  \label{fig:3targets}
\end{figure}

The result from evaluating the KM imaging function \eqref{eq:KM} for
this problem is shown in the left figure of
Fig.~\ref{fig:3targets}. The corresponding result from evaluating the
modified KM imaging function \eqref{eq:rKM} with $\delta = 10^{-2}$ is
shown in the right plot of Fig.~\ref{fig:3targets}. These images show
that the method is capable of identifying the three targets and give
good predictions for their locations. 

\begin{figure}[htb]
  \centering
  \includegraphics[width=0.3\linewidth]{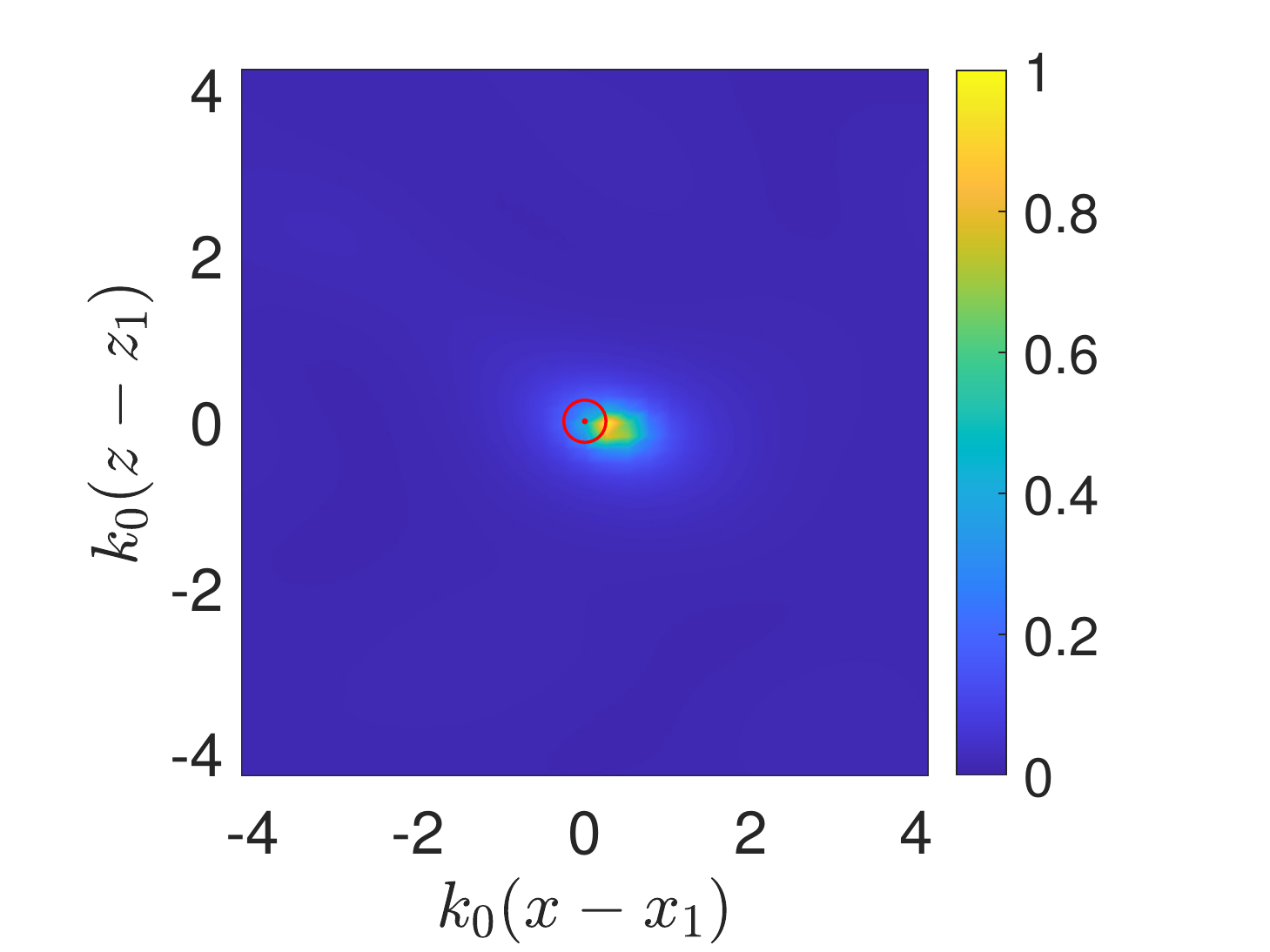}
  \includegraphics[width=0.3\linewidth]{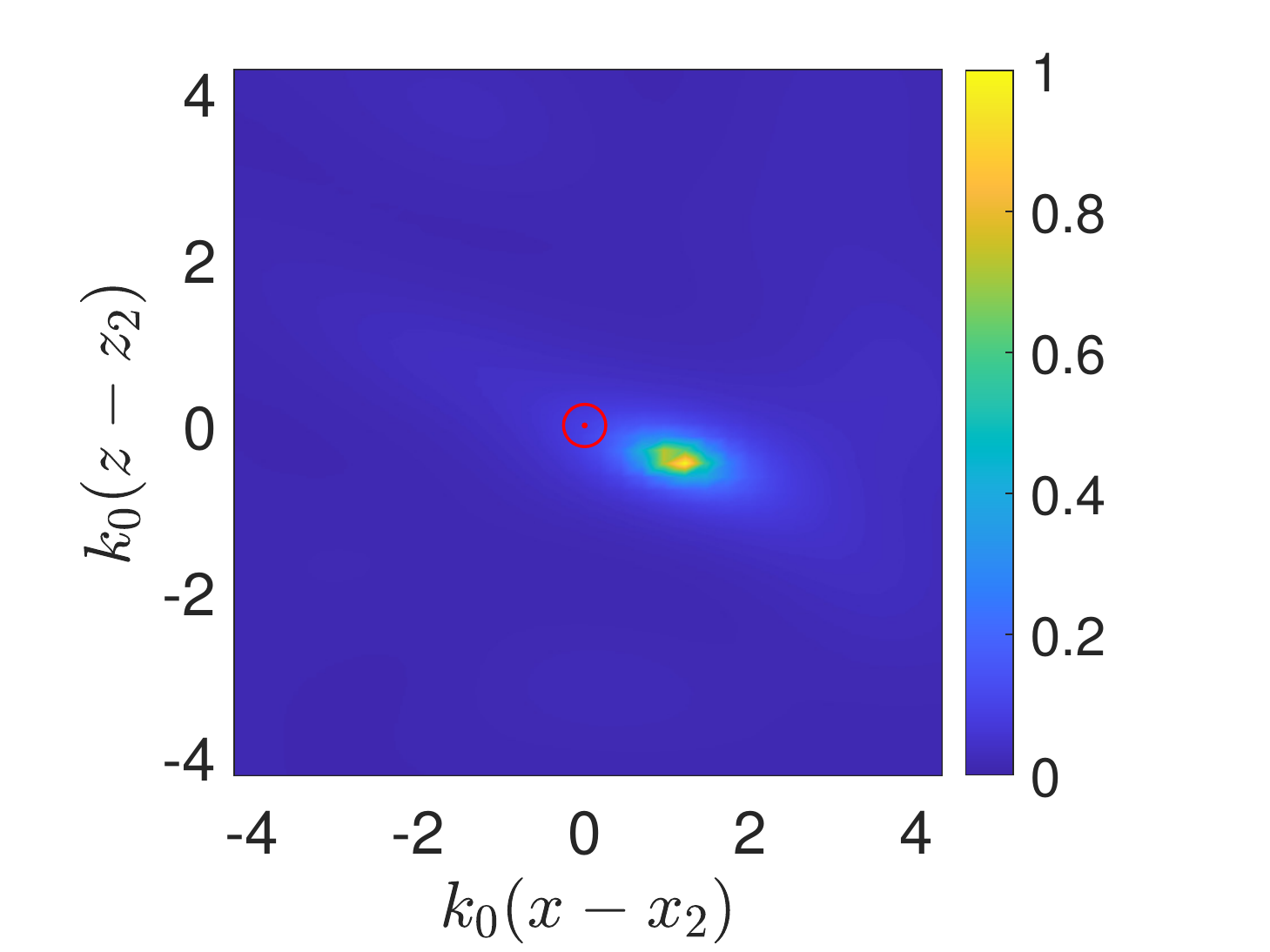}
  \includegraphics[width=0.3\linewidth]{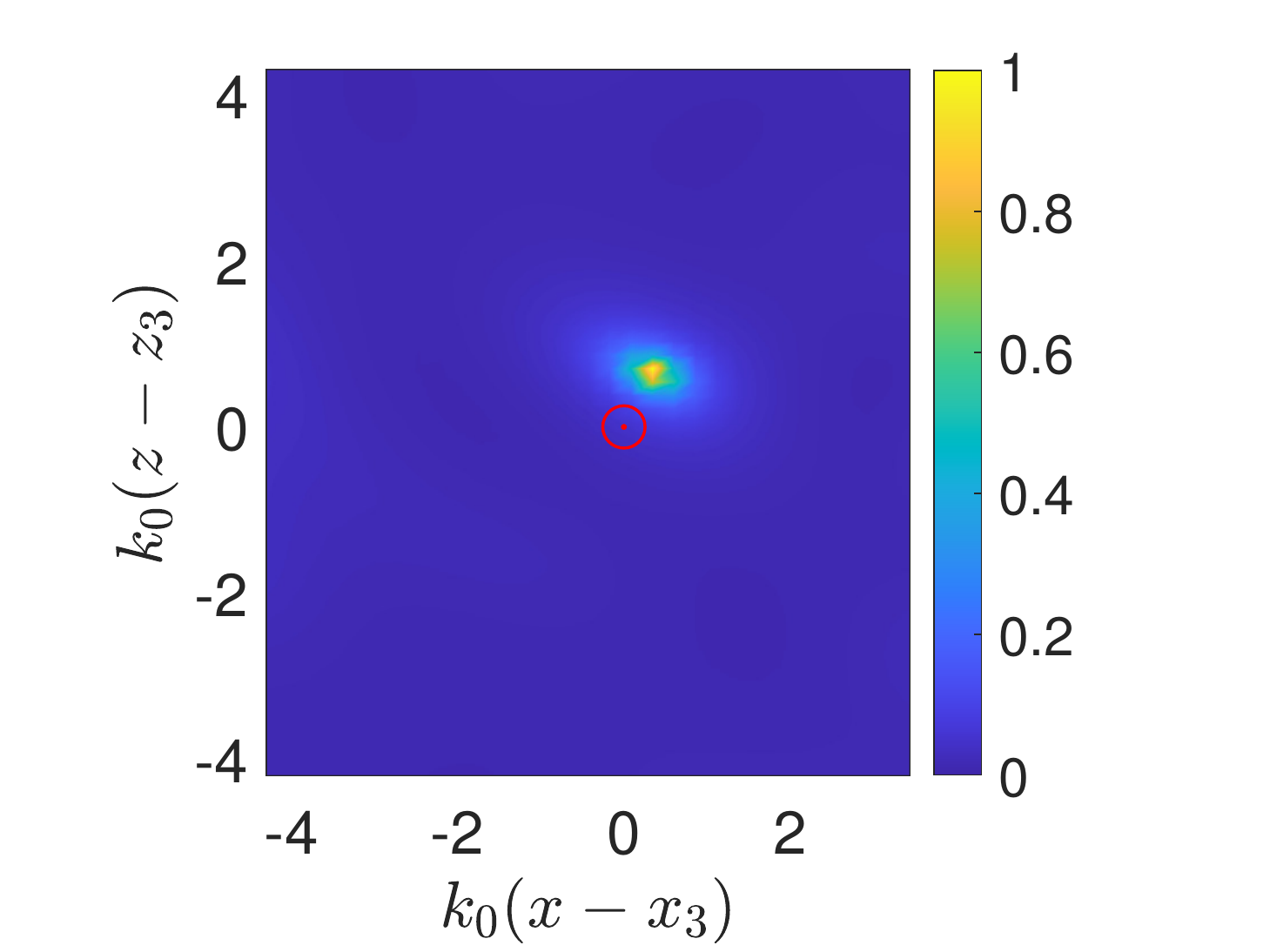}
  \caption{Evaluation of the modified KM imaging function
    \eqref{eq:rKM} with $\delta = 10^{-2}$ in sub-regions centered
    about each target location.}
  \label{fig:3targets-subregions}
\end{figure}

The result from the modified KM method does not show the three targets
equally clearly. In fact, the peak formed near target 2 is the
strongest in the KM image, so the result for the modified KM image
shows target 2 most clearly. This is because the normalization of the
KM image required for evaluating the modified KM image is based on
target 2. As an alternative, we consider
$5\, \text{cm}\, \times\, 5\, \text{cm}$ sub-regions about each of the peaks
of the KM image. Within each of those sub-regions, we normalize the KM
image 
and evaluate the modified KM image with
$\delta = 10^{-2}$. Those results are shown in
Fig.~\ref{fig:3targets-subregions}. Each of those sub-region images is
centered about the corresponding exact target location and scaled by
the central wavenumber $k_{0}$. Even though the predicted target
locations are shifted from the exact target location, these results
show that these shifts are small fractions of the central wavelength.

These results show that this imaging method is capable of identifying
multiple targets. However, there are limitations. The targets cannot
be too close to one another due to the finite resolution of KM
imaging. Moreover, due to absorption in the medium, there are depth
limitations to where targets can be identified. Additionally, when
there are multiple targets at different depths, it is likely that
those targets that are deeper than others may be not be identifiable
in images.
  
\section{Conclusions}
\label{sec:conclusions}

We have discussed synthetic aperture subsurface imaging of point
targets. Here, we have modeled uncertainty about the interface between
the two media with Gaussian-correlated random rough surfaces
characterized by a RMS height and correlation length. The medium above
the interface is uniform and lossless. The medium below the interface
is uniform and lossy. The loss tangent of the medium below the
interface is not known when imaging.

The imaging method involves two steps. First, we attempt to remove
ground bounce signals using principal component analysis. This method
does not require any explicit information about the interface other
than the ground bounce signals is stronger than the scattered
signals. There is no {\it a priori} method to choose the number of
principal components to include in the ground bounce removal
procedure. Instead, we have proposed to determine where the decay of
the singular values changes behavior and use that for the grounce
bounce removal procedure.  Using the resulting matrix after removing
the ground bounce signal, we apply Kirchhoff migration (KM) and our
modification to it that allows for tunably high resolution images of
targets. In our implementation of KM imaging, we compute so-called
illuminations for the problem with a flat interface at the mean
interface height using only the real part of the relative dielectric
permittivity for the medium below that interface, so we completely
neglect the unknown absorption in the medium.

Our numerical results show that despite uncertainty in the interface,
the inexactness of the ground bounce removal procedure, unknown
absorption, and measurement noise, this imaging method is able to
identify and locate targets robustly and accurately. However, there
are limitations to the capabilities of this imaging method. The main
limitation for this imaging method is that targets cannot be too deep
below the interface. Absorption attenuates the scattered power and
depends on the path length of signals. When targets are deep below the
interface, the path length of scattered signals are too large and
attenuation renders those scattered signals undetectable within the
dynamic range of measurements. Additionally, targets cannot be too
closely situated to one another. The KM imaging method is limited in
its resolution. If targets are situated closer than the resolution
capabilities of KM, they cannot be distinguished.

Despite the limitations of this imaging method, we find these results
to be a promising first step toward practical imaging problems. A key
extension of this work will be to incorporate quantitative imaging
methods that will open opportunities for target classification in
addition to identification and location. We have recently developed
methods for recovering the radar cross-section (RCS) for dispersive
point targets when there is no ground bounce
signal~\cite{KT-dispersive}. Recovering the RCS for individual targets
can be used to classify targets by properties related to their size or
material properties when their shape or other geometrical features are
not available for recovery. The challenge with quantitative imaging
methods for this problem will be addressing both the unknown
absorption and uncertain rough interface. As mentioned previously,
absorption will attenuate the power scattered by targets. Moreover, it
will attenuate power non-uniformly over frequency which introduces new
challenges. The uncertainty in the rough interface also affects our
ability to recover quantitative information. Because our method for
removing ground bounce signals from an unknown rough surface is
approximate, it yields errors in the phase which impeded the
recovery of quantitative information. Developing extensions that allow
for quantitative subsurface imaging is the subject of our future work.

\section*{Appendix: Numerical solution of the system of boundary
  integral equations}

\renewcommand{\theequation}{A\arabic{equation}}

\setcounter{equation}{0}

The method that we use to compute realizations of the
Gaussian-correlated rough surface \cite{tsang2004scattering} uses
discrete Fourier transforms, which assumes periodicity over the
interval $[-L/2, L/2]$.  The truncated domain width $L$ is chosen
large enough so that edges do not strongly affect the results. In the
simulations used here we set $L = 4$ m compared to the $1$ m aperture
and $30$ cm wide imaging window.

To compute the numerical solution of \eqref{eq:BIE-top} or
\eqref{eq:BIE-bottom}, we first truncate the integrals to the interval
$-L/2 \le \xi \le L/2$ and then replace those integrals with numerical
quadrature rules. The result of this approximation is a finite
dimensional linear system of equations suitable for numerical
computation. Because the rough surfaces are periodic, we use the
periodic trapezoid rule (composite trapezoid rule for a periodic
domain). However, because the integral operators in \eqref{eq:BIE-top}
and \eqref{eq:BIE-bottom} are weakly singular, we need to make
modifications to the periodic trapezoid rule which we explain below.

We discuss the modification to the periodic trapezoid rule we use for
the integrals,
\begin{equation}
  I_{D}(s) = \int_{-L/2}^{L/2} \frac{\partial
    G(s,h(s);t,h(t))}{\partial n} \sqrt{1 + (h'(t))^{2}}
  U(t) \mathrm{d}t,
  \label{eq:DLP}
\end{equation}
and
\begin{equation}
  I_{S}(s) = \int_{-L/2}^{L/2} G(s,h(s);t,h(t)) V(t) \mathrm{d}t,
  \label{eq:SLP}
\end{equation}
with
\begin{equation*}
  G(s,h(s);t,h(t)) = \frac{\mathrm{i}}{4} H_{0}^{(1)}\left( k
    \sqrt{ ( s - t )^{2} + ( h(s) - h(t) )^{2}} \right).
\end{equation*}
Let $t_{j} = -L/2 + (j-1) \Delta t$ for $j = 1, \dots, M$ denote the
$M$ quadrature points with $\Delta t = L / M$. By applying the
periodic trapezoid rule to \eqref{eq:DLP} and \eqref{eq:SLP} and
evaluating that result on $s = t_{i}$, we obtain
\begin{equation*}
  I_{D}^{M}(t_{i}) = \Delta t \sum_{j = 1}^{M}  \frac{\partial
    G(t_{i},h(t_{i});t_{j},h(t_{j}))}{\partial n} \sqrt{1 +
    (h'(t_{j}))^{2}} U(t_{j}),
\end{equation*}
and
\begin{equation*}
  I_{S}^{M}(t_{i}) = \Delta t \sum_{j = 1}^{M}
  G(t_{i},h(t_{i});t_{j},h(t_{j})) V(t_{j}).
\end{equation*}
Let $A$ be the $M \times M$ matrix whose entries are
\begin{equation}
  a_{ij} = \Delta t \frac{\partial
    G(t_{i},h(t_{i});t_{j},h(t_{j}))}{\partial n} \sqrt{1 +
    (h'(t_{j}))^{2}},
  \label{eq:aij}
\end{equation}
and let $B$ be the $M \times M$ matrix whose entries are
\begin{equation}
  b_{ij} = \Delta t G(t_{i},h(t_{i});t_{j},h(t_{j})).
  \label{eq:bij}
\end{equation}
With these matrices defined, the approximations for the integral
operators given above are matrix-vector products.  The problem with
these results is that the kernels for $I_{D}^{M}$ and $I_{S}^{M}$ are
singular on $t_{j} = t_{i}$, so the diagonal entries of $A$ and $B$
cannot be specified.

The modification to the periodic trapezoid rule we make is to replace
the diagonal entries of $A$ and $B$ by
\begin{equation*}
  a_{ii} = U(t_{i}) \int_{t_{i} - \Delta t/2}^{t_{i} + \Delta t/2}
  \frac{\partial G(t_{i},h(t_{i});t,h(t))}{\partial n} \sqrt{1 +
    (h'(t))^{2}} \mathrm{d}t,
\end{equation*}
and
\begin{equation*}
  b_{ii} = V(t_{i}) \int_{t_{i} -\Delta t/2}^{t_{i} + \Delta t/2}
  G(t_{i},h(t_{i});t,h(t)) \mathrm{d}t.
\end{equation*}
Note that we have assumed that $U(t)$ and $V(t)$ are approximately
constant over this interval thereby allowing us to factor them out
from the integral. Substituting $t = t_{i} + \tau$ and
$\mathrm{d}t = \mathrm{d}\tau$, we obtain
\begin{equation*}
  a_{ii} = U(t_{i}) \int_{- \Delta t/2}^{\Delta t/2} \frac{\partial
    G(t_{i},h(t_{i});t_{i} + \tau,h(t_{i} + \tau))}{\partial n}
  \sqrt{1 + (h'(t_{i} + \tau))^{2}}
  \mathrm{d}\tau,
\end{equation*}
and
\begin{equation*}
  b_{ii} = V(t_{i}) \int_{-\Delta t/2}^{\Delta t/2}
  G(t_{i},h(t_{i});t_{i} + \tau,h(t_{i} + \tau)) \mathrm{d}\tau.
\end{equation*}
Next, we evaluate the expressions involving $G$ and find that
\begin{multline*}
  \frac{\partial G(t_{i},h(t_{i});t_{i} + \tau,h(t_{i} +
    \tau))}{\partial n}
  \sqrt{1 + (h'(t_{i} + \tau))^{2}}\\
  = -\frac{\mathrm{i} k}{4} \left[ h'(t_{i}) \tau - h(t_{i})
    + h(t_{i} + \tau) \right] \frac{H_{1}^{(1)}( k \sqrt{ \tau^{2} + (
      h(t_{i}) - h(t_{i} + \tau) )^{2}} )}{\sqrt{ \tau^{2} + (
      h(t_{i}) - h(t_{i} + \tau) )^{2}}},
\end{multline*}
and
\begin{equation*}
  G(t_{i},h(t_{i});t_{i} + \tau,h(t_{i} + \tau)) =
  \frac{\mathrm{i}}{4} H_{0}^{(1)}( k \sqrt{ \tau^{2} + (
      h(t_{i}) - h(t_{i} + \tau) )^{2}} )
\end{equation*}
Expanding about $\tau = 0$, we find
\begin{equation*}
  \frac{\partial G(t_{i},h(t_{i});t_{i} + \tau,h(t_{i} +
    \tau))}{\partial n}
  \sqrt{1 + (h'(t_{i} + \tau))^{2}} =
  \frac{h''(t_{i})}{4 \pi ( 1 + (h'(t_{i}))^{2} )} + 
  O(\tau^{2}),
\end{equation*}
and
\begin{equation*}
  G(t_{i},h(t_{i});t_{i} + \tau,h(t_{i} + \tau)) =
  \frac{1}{4\pi}\left[ -2 \gamma + \mathrm{i} \pi - 2 \log\left(
      \frac{1}{2} k |\tau| \sqrt{1 + (h'(t_{i}))^{2}} \right)
  \right] + O(\tau^{2}),
\end{equation*}
with $\gamma = 0.5772\dots$ denoting the Euler-Mascheroni constant.
Integrating these expressions over
$-\Delta t/2 \le \tau \le \Delta t/2$, we set
\begin{equation}
  a_{ii} = \frac{\Delta t}{4\pi} \frac{h''(t_{i})}{1 + (h'(t_{i}))^{2}},
  \label{eq:aii}
\end{equation}
and
\begin{equation}
  b_{ii} = \frac{\Delta t}{2\pi} \left[ 1 - \gamma + \mathrm{i} \frac{\pi}{2}
    - \log\left( \frac{1}{4} k \Delta t \sqrt{ 1 + (h'(t_{i}))^{2}}
    \right) \right].
  \label{eq:bii}
\end{equation}

Thus, to form the matrix $A$, we evaluate \eqref{eq:aij} for all $i
\neq j$ and \eqref{eq:aii} for $i = j$. Similarly, to form the matrix
$B$, we evaluate \eqref{eq:bij} for all $i \neq j$ and \eqref{eq:bii}
for $i = j$. With these matrices, we seek the vectors of unknowns,
$\mathbf{u} = ( U(t_{1}), \dots, U(t_{M}) )$ and 
$\mathbf{v} = ( V(t_{1}), \dots, V(t_{M}) )$ through solution of the
block system of equations,
\begin{equation*}
  \begin{bmatrix} \frac{1}{2} I - A_{0} & B_{0} \\ \frac{1}{2} I +
    A_{1} & -B_{1} \end{bmatrix}
  \begin{bmatrix} \mathbf{u}\\ \mathbf{v} \end{bmatrix}
  = \begin{bmatrix} \mathbf{f}_{0} \\ \mathbf{f}_{1} \end{bmatrix}.
\end{equation*}
Here $I$ is the identity matrix, $A_{0}$ and $B_{0}$ correspond to
evaluation of the $A$ and $B$ matrices with wavenumber $k_{0}$ and
$A_{1}$ and $B_{1}$ correspond to evaluation of the $A$ and $B$
matrices with wavenumber
$k_{1} = k_{0} \sqrt{\epsilon_{r} (1 + \mathrm{i} \beta)}$. The
right-hand side block vectors contain the evaluation of the source
above the interface $\mathbf{f}_{0}$ and below the interface
$\mathbf{f}_{1}$ on the set of interface points $(t_{j}, h(t_{j}))$
for $j = 1, \dots, M$.

\section*{Acknowledgments}

The authors acknowledge support by the Air Force Office of Scientific
Research (FA9550-21-1-0196). A.~D.~Kim also acknowledges support by
the National Science Foundation (DMS-1840265).

\section*{Data Availability Statement} 

The data and numerical methods used in this study are available at
Zenodo via\\ https://doi.org/10.5281/zenodo.7754256


\bibliographystyle{apacite}

\end{document}